%% file: main.tex
\documentclass[9pt,sigconf,letterpaper,nonacm]{acmart}
\usepackage{xurl}
\usepackage{hyperref}
\usepackage{tikz}
\usepackage{amsmath}
\usepackage{makecell}
\usepackage{url,color,soul,xspace,enumitem}
\usepackage{float}
\usepackage{graphicx}
\usepackage{caption}
\usepackage{subcaption}
\usepackage{comment}
\usepackage{xspace}
\usepackage{multicol}
\usepackage{cleveref}
\usepackage{multirow}
\usepackage{enumitem}
\usepackage{xcolor}
\usepackage{pifont} 
\usepackage{tcolorbox}
\usepackage{tikz}
\usepackage{amsfonts}
\usepackage{amsmath}
\usepackage[inline,draft,nomargin,index]{fixme}
\usepackage{booktabs}
\usepackage{tabu}
\usepackage{tabularx}

\usepackage[htt]{hyphenat}

\renewcommand{\paragraph}[1]{\smallskip\noindent{\bf #1}~~~}

\newenvironment{packed_itemize}{
\begin{itemize}[leftmargin=3mm,nolistsep,noitemsep]
  \setlength{\itemsep}{0pt}
  \setlength{\parskip}{0pt}
  \setlength{\parsep}{0pt}
  \setlength{\topsep}{0pt}
}{\end{itemize}}

\newenvironment{packed_enumerate}{
    \begin{enumerate}[leftmargin=6mm,label=(\roman*),nolistsep,noitemsep]
  \setlength{\itemsep}{0pt}
  \setlength{\parskip}{0pt}
  \setlength{\parsep}{0pt}
  \setlength{\topsep}{0pt}
}{\end{enumerate}}

\newcommand{\www}{\texttt{www}\xspace}
\newcommand{\cf}{\textit{Cloudflare}\xspace}
\newcommand{\adb}{\texttt{AD}\xspace}

\newcommand{\svcb}{\texttt{SVCB}\xspace}
\newcommand{\https}{\texttt{HTTPS}\xspace}
\newcommand{\arec}{\texttt{A}\xspace}
\newcommand{\aaaa}{\texttt{AAAA}\xspace}
\newcommand{\ns}{\texttt{NS}\xspace}
\newcommand{\soa}{\texttt{SOA}\xspace}
\newcommand{\cname}{\texttt{CNAME}\xspace}
\newcommand{\dname}{\texttt{DNAME}\xspace}
\newcommand{\dnskey}{\texttt{DNSKEY}\xspace}
\newcommand{\ds}{\texttt{DS}\xspace}
\newcommand{\rrsig}{\texttt{RRSIG}\xspace}

\newcommand{\amode}{\texttt{AliasMode}\xspace}
\newcommand{\smode}{\texttt{ServiceMode}\xspace}
\newcommand{\sprty}{\texttt{SvcPriority}\xspace}
\newcommand{\tname}{\texttt{TargetName}\xspace}
\newcommand{\sparam}{\texttt{SvcParams}\xspace}
\newcommand{\mdtr}{\texttt{mandatory}\xspace}
\newcommand{\alpn}{\texttt{alpn}\xspace}
\newcommand{\noalpn}{\texttt{no-default-alpn}\xspace}
\newcommand{\port}{\texttt{port}\xspace}
\newcommand{\ipfh}{\texttt{ipv4hint}\xspace}
\newcommand{\ipsh}{\texttt{ipv6hint}\xspace}
\newcommand{\ech}{\texttt{ech}\xspace}

\newcommand{\chello}{\textit{ClientHello}\xspace}
\newcommand{\shmode}{\texttt{Shared Mode}\xspace}
\newcommand{\spmode}{\texttt{Split Mode}\xspace}

\newcommand{\insecure}{\texttt{insecure}\xspace}
\newcommand{\bogus}{\texttt{bogus}\xspace}

\newcommand{\halfcirc}[1][0.7ex]{%
  \begin{tikzpicture}
  \draw[fill] (0,0)-- (90:#1) arc (90:270:#1) -- cycle ;
  \draw (0,0) circle (#1);
  \end{tikzpicture}}
\newcommand{\fullcirc}[1][0.7ex]{\tikz\fill (0,0) circle (#1);}
\newcommand{\emptycirc}[1][0.7ex]{\tikz\draw (0,0) circle (#1);}

\begin{document}

\newcommand{\PaperTitle}{Exploring the Ecosystem of DNS HTTPS Resource Records:\\An End-to-End Perspective}
\newcommand{\PaperNumber}{XXX}
\title{\PaperTitle}

\author{Hongying Dong\textsuperscript{*}}
\affiliation{%
\institution{University of Virginia}
\city{Charlottesville}
\state{Virginia}
\country{USA}
}
\email{hd7gr@virginia.edu} 

\author{Yizhe Zhang\textsuperscript{*}}
\affiliation{%
\institution{University of Virginia}
\city{Charlottesville}
\state{Virginia}
\country{USA}
}
\email{yz6me@virginia.edu}

\author{Hyeonmin Lee}
\affiliation{%
\institution{University of Virginia}
\city{Charlottesville}
\state{Virginia}
\country{USA}
}
\email{frv9vh@virginia.edu}

\author{Shumon Huque}
\affiliation{%
\institution{Salesforce}
\city{McLean}
\state{Virginia}
\country{USA}
}
\email{shuque@gmail.com}

\author{Yixin Sun}
\affiliation{%
\institution{University of Virginia}
\city{Charlottesville}
\state{Virginia}
\country{USA}
}
\email{ys3kz@virginia.edu}

\renewcommand{\shortauthors}{Hongying Dong, Yizhe Zhang, Hyeonmin Lee, Shumon Huque, \& Yixin Sun}

\thanks{\textsuperscript{*} Both Hongying Dong and Yizhe Zhang contributed equally to this work.}

\begin{abstract}
\input{sections/0_abstract}
\end{abstract}


\maketitle

\input{sections/1_intro}
\input{sections/2_background}
\input{sections/3_dataset}

\input{sections/4_server}

\input{sections/5_client}

\input{sections/7_relatedwork}

\input{sections/8_1_discussion}

\input{sections/8_2_conclusion}

\begin{acks}
We thank anonymous reviewers for their insightful and constructive suggestions and feedback.
This work is supported by National Science Foundation CNS-2154962 and CNS-2319421, and the Commonwealth Cyber Initiative.
\end{acks}


\appendix
\input{sections/appendix}
\end{document}

%% file: sections/0_abstract.tex
The DNS \https resource record is a new DNS record type designed for the delivery of configuration information and parameters required to initiate connections to \https network services. 
In addition, it is a key enabler for TLS Encrypted ClientHello (ECH) by providing the cryptographic keying material needed to encrypt the initial exchange. To understand the adoption of this new DNS \https record, we perform a longitudinal study on the server-side deployment of DNS \https for Tranco top million domains, as well as an analysis of the client-side support for DNS \https through snapshots from major browsers. To the best of our knowledge, our work is the first longitudinal study on DNS \https server deployment, and the first known study on client-side support for DNS \https. 
Despite the rapidly growing trend of DNS \https adoption, our study highlights challenges and concerns in the deployment by both servers and clients, such as the complexity in properly maintaining \https records and connection failure in browsers when the \https record is not properly configured. 

%% file: sections/1_intro.tex
\section{Introduction}
\label{sec:intro}



Transport Layer Security (TLS) plays a pivotal role in securing the Internet. 
Notably, Hypertext Transfer Protocol Secure (HTTPS) is an extension of Hypertext Transfer Protocol (HTTP) that employs TLS to protect web communications (e.g., communications between a web browser and a website).

However, upgrading a connection from HTTP to HTTPS typically incurs additional latency.
Since a browser initially does not have knowledge of whether a website supports HTTPS, it typically attempts to first send a plaintext HTTP request.\footnote{This occurs if a user enters a domain name (e.g., a.com) in the browser’s address bar without adding HTTPS prefix (e.g., https://a.com).}
The connection is only upgraded to HTTPS if the website responds with an HTTPS redirect.
The plaintext nature of the HTTP to HTTPS redirection process presents a potential target for man-in-the-middle attackers to block or redirect clients to their own (malicious) HTTPS site.
Although the HTTP Strict Transport Security (HSTS)~\cite{rfc6797} policy and Alt-Svc header~\cite{rfc7838} can mitigate these issues to some degree, they do not negate the need for the browser's first HTTP request and the HTTPS redirect. 
Additionally, while pre-configured HSTS preload lists can be used by browsers to unconditionally connect to websites on the list using HTTPS, the process of populating such lists is manual and does not cover the vast majority of websites.


The recently standardized \svcb and \https DNS Resource Records (RR)~\cite{rfc9460} offer promising approaches to address these concerns. 
\svcb records provide clients with comprehensive information need-ed to access a service, including supported protocols, port numbers, and IP addresses by directly storing the information in the DNS record.
In particular, the \https record, a variation of \svcb tailored to the HTTPS protocol, informs clients about a website's HTTPS support, along with additional details such as supported HTTP versions. 
Therefore, a client can obtain all necessary information for accessing a website through a single \https DNS query, thereby enabling it to directly establish a TLS session using this information.
In contrast to the \cname record, which aliases the entire domain and thereby excludes the inclusion of other record types, the \https record supports coexistence with various record types and offers enhanced connection capabilities. Unlike the \texttt{DNAME} record that redirects an entire DNS subtree to another subtree, \https records can redirect web traffic specifically at the \texttt{DNAME} owner while permitting distinct redirection policies for subdomains. Additionally, \https records may be employed within the \texttt{DNAME} subtree itself.
Furthermore, another important aspect that \https records could facilitate is the conveyance of TLS Encrypted Client Hello (ECH)~\cite{ietf-tls-esni-17} information, which encrypts the \chello message during the TLS handshake.
Although ECH is not yet standardized, its integration with \https records could enhance the privacy of TLS connections.

Given the performance and security benefits provided by the \https record, it has been adopted by large cloud providers such as Cloudflare~\cite{cloudflare-https} and Akamai~\cite{akamai-https}, even before its standardization in November 2023.
Furthermore, popular web browsers (e.g., Chrome~\cite{chrome-https}, Firefox~\cite{firefox-https}, and Safari~\cite{safari-https}) and DNS software (e.g., BIND 9~\cite{bind9-https}, PowerDNS~\cite{powerdns-https}, and Knot DNS~\cite{knotdns-https}) support \https records.


Nevertheless, the understanding of the current landscape of \https record is still limited.
While a preliminary study~\cite{zirngibl2023first} provides brief statistics on server-side \https records through one snapshot, it does not perform a longitudinal analysis or investigate client-side support.
This research gap limits the understanding of the effectiveness and challenges of both server-side and client-side \https record deployment.

In this paper, we present an \textit{extensive} study of the DNS \https ecosystem, by encompassing both server-side and client-side deployments.
For server-side analysis, we take daily snapshots of \https records from May 2023 to Mar 2024 for domains in the Tranco list~\cite{le2019tranco}, which ranks the top 1 million domains based on their popularity across various lists.
Specifically, we scan the primary apex domains (e.g., \texttt{a.com}) and their corresponding \www subdomains (e.g., \texttt{www.a.com}).
We also collect other relevant data such as the \ns and \soa records (from Aug 2023 to Mar 2024) and WHOIS information of name servers (from Oct 2023 to Mar 2024) to facilitate our analysis. 

To examine client-side behavior, we investigate how popular web browsers support \https records, including their failover mechanisms, by configuring our own DNS server with \https records and performing active experiments with the browsers.
To the best of our knowledge, there is no existing research on browsers' support of \https records.

\paragraph{Key contributions.} We present the first longitudinal study on the deployment of DNS \https records by top domains and perform extensive testing of browser behavior in handling DNS \https requests. Our measurements allow us to gain a deeper understanding of the DNS \https ecosystem, and highlight potential obstacles and concerns that can help inform future deployment. 
Our key findings are:
\begin{packed_itemize}
    \item Despite its recent standardization, over 20\% of Tranco top 1M domains have DNS \https records and major browsers utilize \https records in establishing connections. Notably, a major contributing factor is \textit{Cloudflare}'s default \https configuration, which accounts for over 70\% of domains with \https records. 
    \item The adoption of \https records inevitably incurs complexity and  overhead in properly maintaining the validity of records to avoid connection failure. 
    For example, the frequent ECH key rotation every 1 to 2 hours requires the server to properly implement retry mechanisms~\cite{ietf-tls-esni-17} to avoid breaking the connection due to expired keys. 
    \item Connection failures from major browsers occur in various \https misconfigurations due to the lack of proper failover mechanisms, exacerbating the challenge of utilizing \https records for both servers and clients. Furthermore, the lack of support for ECH \spmode leads to such failures across all major browsers, prohibiting clients from establishing connections to the server even when ECH is correctly configured by the server.
\end{packed_itemize}

\paragraph{Artifact availability.} 
We provide full availability, including our dataset and code to reproduce our results at \url{https://github.com/yzzhn/imc2024dnshttps}. 
In the long run, we plan to maintain a longstanding framework that continuously collects and releases \https data periodically. 

%% file: sections/2_background.tex
\section{Background}
\label{sec:background}


We briefly discuss background on DNS \https records and TLS Encrypted Client Hello (ECH) extension. Further details on DNS record types can be found in Appendix~\ref{appendix:background}.



\paragraph{DNS \https records.}
The \https (and \svcb) record~\cite{rfc9460} is designed to offer alternative endpoints for a service, along with parameters associated with each endpoint, within a single DNS record.
The \https record signals the use of HTTPS (rather than HTTP) for the specified host.
Additionally, it can coexist with other record types, thus enabling name redirection at both zone apexes and any arbitrary location within a zone, a feature not supported by the \cname record.

\begin{figure}[h!]
\vspace{-1mm}
\centering
\setlength{\abovecaptionskip}{8pt}
\setlength{\belowcaptionskip}{-10pt}
\resizebox{\linewidth}{!}{
\setlength{\tabcolsep}{2pt}
\renewcommand{\arraystretch}{1.1}
\texttt{
\begin{tabular}{|cccccll|}
    \hline
    a.com. & 300 & IN & HTTPS & 0 & \multicolumn{2}{l|}{b.com.}  \\
    c.com. & 300 & IN & HTTPS & 1 & . & alpn=h3 ipv4hint=1.2.3.4 \\
    \hline
\end{tabular}
}
}
\caption{An example of \https records. }
\label{fig:example_record}
\end{figure}

\Cref{fig:example_record} illustrates example \https records.
Each \https record consists of two required fields and one optional field:
\begin{packed_itemize}
    \item{\sprty}: the priority of the record (lower values preferred). A value of zero indicates \amode, aliasing a domain to the target domain (specified in \tname). Other values indicate \smode, which provides information specific to a service endpoint.
    \item{\tname}: a domain name, which can be either the alias target in \amode or the alternative endpoint in \smode. If \smode specifies the value ``.'', the owner name of this record has to be used.
    \item{\sparam} (optional): Utilized only in \smode, a list of key-value pairs are included to provide details about the endpoint (in \tname). 
    The current specification defines seven parameter keys (\port, \alpn, \noalpn, \ipfh, \ipsh, \ech, \mdtr).
    The \port parameter specifies additional ports supported by the endpoint, while \alpn (Application-Layer Protocol Negotiation) specifies additional application protocols; by default, an \https record indicates HTTP/1.1 support.
    The \noalpn key is used if the endpoint doesn't support the default protocol. \ipfh/\ipsh suggest IPv4/IPv6 addresses for reaching the endpoint.
    The \ech parameter can be used to include the Encrypted Client Hello (ECH) information.
    The \mdtr parameter specifies mandatory keys that must be used to connect to the endpoint.
\end{packed_itemize}

One notable functionality of the \https record is its ability to publish ECH information.
In the current specification~\cite{rfc9460}, \ech is a reserved \texttt{SvcParam}, as the ECH specification has not yet been standardized.


\textbf{Comparison with DNS \cname and \dname records:} The DNS \https record is expected to offer an improved replacement for the \cname record commonly used today to redirect websites to a third party or alternate location. The \cname record is more general purpose in nature since it completely aliases one domain name to another location. Since the \cname aliases the entire domain name (including all the record types at that name), no other record types can exist at the origin domain name. This precludes its use as a web redirection mechanism at the apex of a DNS zone (since the apex necessarily includes other record types like \texttt{NS} and \texttt{SOA}). By contrast, the \https record is application- and type-specific, which can coexist with record types, and offers additional connection level capability discovery and an extensible framework for new parameters. The \svcb record (the more generalized form of \https records) is more similar to the existing DNS \texttt{SRV} record~\cite{rfc2782}, where the record name identifies the service (and optionally other parameters like port number) via additional labels prepended to the domain name, but offers all the additional connection level capability indications that the \texttt{SRV} record does not.

The \texttt{DNAME} record~\cite{rfc6672} redirects an entire DNS subtree underneath the owner domain name to another DNS subtree, and is not directly comparable to the \https record, although they can co-exist in various ways. For example, an \https record could conceivably redirect web traffic at the owner of the \texttt{DNAME}, while subdomains of the \texttt{DNAME} are redirected elsewhere. 
\https records could be placed at domains names in the redirected \texttt{DNAME} subtree.

\paragraph{TLS Encrypted Client Hello (ECH).}
ECH~\cite{ietf-tls-esni-17} is a TLS extension that allows a client to encrypt its initial \chello message in the TLS handshake.
Normally, the \chello message is sent in plaintext, 
revealing information such as the server's domain name (SNI).\footnote{Although TLS 1.3 encrypts most handshake messages, the \chello message remains unencrypted.} 
To enable ECH, a domain needs to publish key information (e.g., its public key for encrypting the \chello message).
The \https record provides a way for this publication, allowing a domain to include its key information as the \ech parameter.
Therefore, a client can retrieve the \https record and use the \ech parameter to encrypt its \chello message.
%
Currently, major browsers like Chrome~\cite{chrome-ech} and Firefox~\cite{firefox-ech} have implemented ECH.

\paragraph{Tranco list.}
The Tranco List~\cite{le2019tranco} offers a thorough and up-to-date ranking of the internet's most visited websites by aggregating data from a range of sources. This list synthesizes diverse traffic measurements to produce a unified ranking system that accurately reflects the relative popularity of websites. Its integration of multiple traffic metrics enhances both the precision and reliability of the rankings.
In this study, we use the Tranco list as the basis for investigating trends in \https deployment across popular domains. 

%% file: sections/3_dataset.tex
\section{Research Overview}
\label{sec:dataset}



We provide a brief overview of our research goals and agenda.

\paragraph{(1) Server-side \https record deployment}
\begin{itemize}[leftmargin=6mm,topsep=3pt]
  \setlength{\itemsep}{4pt}
  \setlength{\parskip}{0pt}
  \setlength{\parsep}{0pt}
  \setlength{\topsep}{2pt}
  
    \item \textbf{Goal}: Our aim is to analyze the deployment trends of \https records, as well as their characteristics.
    \item \textbf{Methodology}: We measure the deployment of \https records, focusing on the popular domains within the Tranco top one million domain list. 
    Our analysis includes details on the \https record configurations, along with the support of other security protocols such as ECH and DNSSEC in conjunction with \https records.
    \item \textbf{Datasets}: We collect \https records as well as other DNS records, including \arec, \aaaa, \soa, and \ns records, every day from the Tranco 1 million domains. We also scan the IP addresses (\arec/\aaaa records) of name servers used by domains that deploy \https records. Additionally, we utilize the WHOIS database and perform DNSSEC record validation to further analyze the management of \https records from these domains. The details of the server-side datasets are described in \Cref{subsec:server_dataset}.
\end{itemize}

\paragraph{(2) Client-side \https record support}
\begin{itemize}[leftmargin=6mm,topsep=3pt]
  \setlength{\itemsep}{4pt}
  \setlength{\parskip}{0pt}
  \setlength{\parsep}{0pt}
  \setlength{\topsep}{2pt}
  
    \item \textbf{Goal}: We aim to examine support of \https records and identify associated behaviors in popular web browsers. 
    \item \textbf{Methodology}: We focus on the top four browsers: Chrome, Safari, Edge, and Firefox. We analyze whether these browsers (i) perform \https records lookup, (ii) utilize the information in the \https records, and (iii) how they respond to incorrect/inaccurate \https records. We set up our own DNS server and configure \https records to perform controlled experiments. The setup and methodology are detailed in \Cref{sec:client}.
\end{itemize}



We provide insights into the adoption of DNS \https records through quantitative server-side analysis and examine the impact of web browser functionalities on \https record support via the client-side evaluation. These approaches together present a comprehensive view of the \https record ecosystem and highlight the challenges in the current deployment.

%% file: sections/4_server.tex
\section{Server-side \https RR Deployment}
\label{sec:serv}

We first describe our data collection. We then delve into the details of server-side \https RR configurations.

\subsection{Datasets}
\label{subsec:server_dataset}


\vspace{-1.5mm}
\paragraph{Scanning framework.}
Our scanning framework retrieves the top 1 million domain list from the Tranco~\cite{le2019tranco} daily, and performs daily scans of \https records along with other DNS records, as shown in \Cref{tab:server_dataset}. 
The Tranco list does not differentiate between apex domains (i.e., \texttt{a.com}) and \www subdomains (i.e., \texttt{www.a.com}), and includes both types.
We preprocess by retrieving the apex domains of all top million domains, and generate \www subdomains by adding \www prefix to the apex domains.
We distinctly consider apex domains and \www subdomains, where the latter have more specific usage for the web.
This results in two lists: (1) a one-million list of apex domains, and (2) a corresponding one-million list of \www subdomains.
\input{tables/tab-dataset}

Next, to scan DNS records of the domain lists, we implement a scanner by utilizing the \textit{dnspython}~\cite{dnspython} library. 
For each domain in the list, we initiate a DNS \https query to two widely recognized public DNS resolvers: \textit{Google} (8.8.8.8) as the main resolver and \textit{Cloudflare} (1.1.1.1) as the backup resolver. 
If a domain produces a \cname response, our inquiry extends to sending an \https query to the domain in the resolved \cname.\footnote{Note that \cname records at the apex of a zone are technically not allowed~\cite{rfc1912}, although some misconfigured DNS servers allow them to be installed.}
For \https records, we also collect the corresponding \rrsig records if provided, along with the \https record.
Additionally, we gather information included in DNS responses of \https records, such as the \texttt{Authenticated Data (AD)}~\cite{rfc3655} bit, which indicates that the response is authenticated with DNSSEC.
If either the domain or the one in the \cname has an \https record, we proceed to perform additional queries for the domain, including \arec (IPv4 address), \aaaa (IPv6 address), \soa, and \ns records lookup.

Additionally, we perform daily scans of \arec/\aaaa records for name servers used by domains that publish \https records to analyze the distribution and trend.
The name servers queried are obtained from the \ns records collected through the daily apex and \www domain scanning. Additionally, we perform WHOIS lookups for IP addresses found in the \arec/\aaaa records of name servers, complemented by manual reviews (as described in~\cref{subsubsec:ns}), to ascertain the registered owner of these IPs. This information is utilized to further analyze the operators of each name server.
Our dataset on name servers covers the period from October 11th, 2023 to March 31st, 2024.
Further discussion on ethics can be found in Appendix~\ref{appendix:ethics}.

\paragraph{Analyzing domains in the Tranco list.}
Starting from August 1st, 2023, Tranco changed its data sources by phasing out Alexa ranking for Chrome User Experience Report and \textit{Cloudflare} Radar data~\cite{tranco-method}, impacting the domain list composition.
Therefore, our analysis is split into two phases, before and after August 1st, 2023, to account for the major change in the Tranco top 1M domain list.

Furthermore, since the Tranco list is updated \textit{daily}, the composition of domains within the list may change each day.
Consequently, relatively popular domains (e.g., those with higher rankings) are likely to be consistently included in the list, while domains with lower rankings may experience fluctuations in their presence on the list. 
Detailed observations of Tranco domain rankings can be found in \Cref{appendix:tranco}.



Therefore, our subsequent analysis distinctively considers two sets of domains, defined as follows: 

\begin{packed_enumerate}
    \item \textbf{Dynamic Tranco top 1M domains}: this set of domains includes the entire 1 million domains in the daily Tranco list. Analyzing this set allows us to observe general trends in \https record deployment of top domains. 
    
    \item \textbf{Overlapping domains}: this set of domains consists of domains that appear in the Tranco list \emph{every day} during our entire measurement period. Due to the changes in Tranco's data source, we further divide our analysis period into two parts. The first part, spanning from May 8th to July 31st, 2023, precedes Tranco's source change and includes 634,810 unique overlapping domains. The second part, from August 1st, 2023, to March 31st, 2024, follows the source change and includes 684,292 unique overlapping domains. By analyzing these overlapping domains, we aim to understand the behavior within a stable (and potentially more popular) set of domains.
\end{packed_enumerate}


\subsection{\https RR Adoption}
\label{subsec:httpsrr}
We analyze the adoption of \https records from the perspective of domains and name servers.

\subsubsection{Adoption rates}
\label{subsubsec:httpsrr}

\begin{figure}[!ht]
    \centering
    \begin{subfigure}[b]{.99\columnwidth}
        \includegraphics[width=.99\columnwidth]{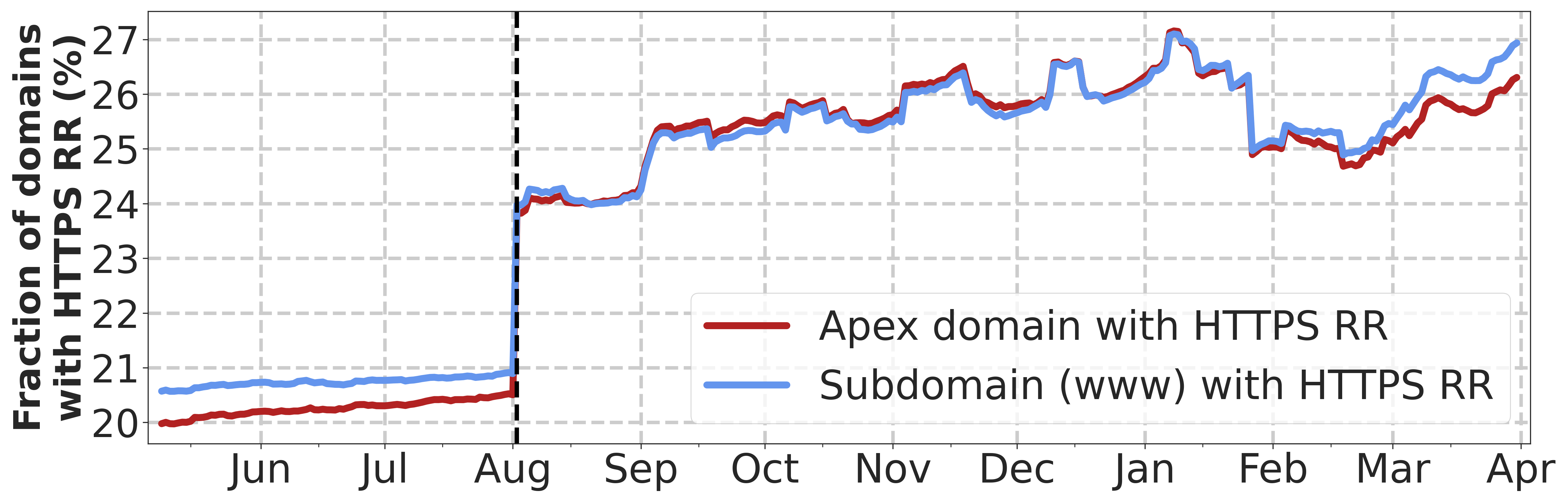}
        \caption{Dynamic Tranco top 1M apex and \www subdomains.}
        \label{fig:alldom-adopt}
    \end{subfigure}
    \begin{subfigure}[b]{.99\columnwidth}
        \setlength{\belowcaptionskip}{-8pt}
        \includegraphics[width=.99\columnwidth]{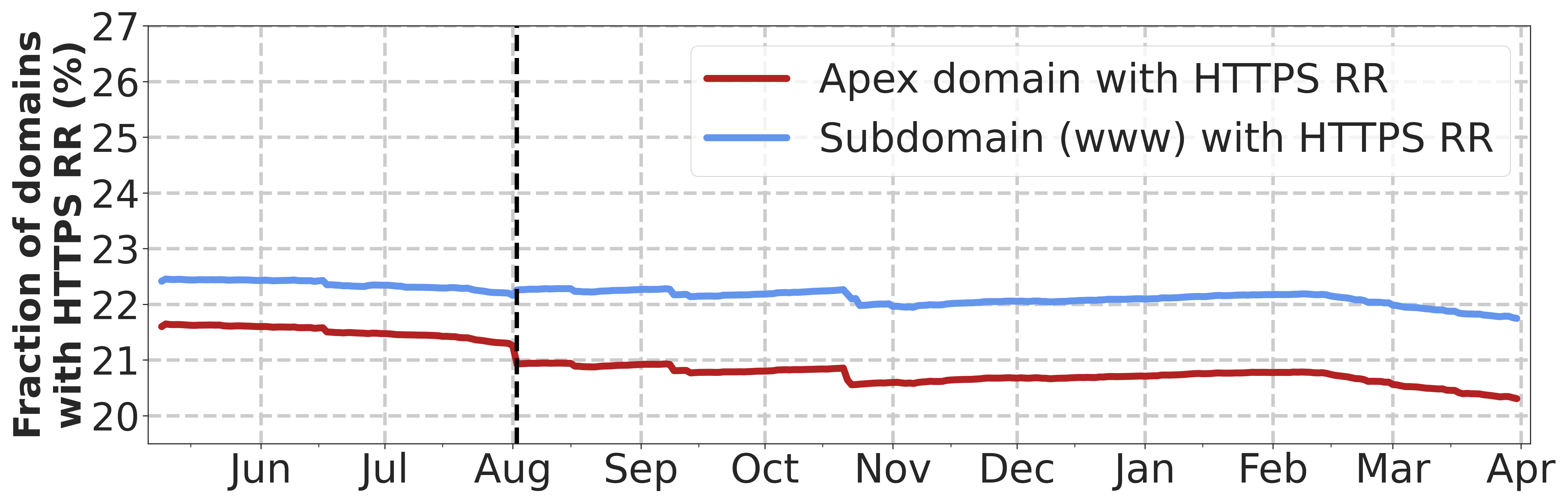}
        \caption{Overlapping apex and \www subdomains.}
        \label{fig:overlap-adopt}
    \end{subfigure}
    \setlength{\belowcaptionskip}{-8pt}
    \caption{Percentages of apex/\www domains that publish \https records. Vertical dashed line (on August 1st, 2023) denotes the source change of the Tranco list.}
    \label{fig:httpsrr-adopt}
\vspace{2mm}
\end{figure}

We start our analysis by examining the deployment of \https records by domains in the Tranco list.
\Cref{fig:httpsrr-adopt} shows the percentage of apex domains 
and their corresponding \www subdomains that publish \https records, from May 8th, 2023, to March 31st, 2024, ranging from 20\% to 27\%. 

We notice that the \https adoption rates exhibit distinct trends between the dynamic Tranco 1M domains (\Cref{fig:alldom-adopt}) and the overlapping domains (\Cref{fig:overlap-adopt}).
While an overall continuously increasing trend is observed in adopting \https RR for the dynamic Tranco top domains, overlapping apex domains and corresponding \www subdomains show a relatively stable ratio.
This stability is disrupted only by a decrease in apex domains and a minor increase in \www subdomains, which occur when Tranco updates its source feeds. This is followed by a gradual decline, likely attributable to changes in name servers (detailed in \Cref{subsubsec:ns}).
Thus, the increase in \https records deployment is more attributed to daily domains that are \emph{not} always in the Tranco top 1M domain list (i.e., non-overlapping domains).







\begin{table}[!ht]
\centering
\resizebox{0.96\columnwidth}{!}{%
\begin{tabular}{ c|cc|cc }
      \multirow{3}{*}{\shortstack[c]{\bf Name Server (NS)  \\ \bf Category}} & \multicolumn{2}{c|}{\bf Dynamic} & \multicolumn{2}{c}{\multirow{2}{*}{\bf Overlapping}} \\
      & \multicolumn{2}{c|}{\bf Tranco top 1M} & & \\
      & Mean (\%) & Std.& Mean (\%) & Std. \\
     \hline
     \makecell[l]{Full \textit{Cloudflare} NS} & 99.89 & 0.03 & 99.87 & 0.04\\       
     \makecell[l]{None \textit{Cloudflare} NS} & 0.11 & 0.03 & 0.13 & 0.04 \\
     \makecell[l]{Partial \textit{Cloudflare} NS} & $<$ 0.01 & - & $<$ 0.01 & - \\  
     
\end{tabular}
}
\caption{The average and standard deviation of the percentage of apex domains (with \https records) served by \textit{Cloudflare} vs non-\textit{Cloudflare} name servers.}
\label{tab:nsinfoalldom}
\vspace{-10mm}
\end{table}

\subsubsection{Name servers}
\label{subsubsec:ns}

Considering that the name server providers could have a considerable impact on the adoption of \https RR, we now take a closer look at 
the distribution of name servers utilized by apex domains that adopt the \https RR. 
We utilize additional data, including \arec/\aaaa records of name servers and WHOIS information
for the corresponding IP addresses (detailed in~\Cref{tab:server_dataset})
to pinpoint the host organization of each name server.\footnote{We use \textit{ipwhois}~\cite{ipwhois} to parse WHOIS information.}
Additionally, we conducted a supplementary manual review of DNS providers. This involved a thorough analysis of their documentation to exclude instances where domain owners utilize cloud service providers for hosting while operating their own name servers (e.g., Amazon AWS). Such cases may result in AS information that misleadingly attributes the name servers to the cloud service provider.
Note that a limitation with using WHOIS occurs when customers use their own IP addresses with cloud service providers (a practice known as BYOIP, or Bring Your Own IP). In such cases, the WHOIS information may reflect the original owner rather than the cloud service provider. This limitation likely affects the tail distribution of name server providers, but not top providers such as Google and GoDaddy.

\begin{figure}[ht]
 \centering
    \setlength{\abovecaptionskip}{0pt}
    \setlength{\belowcaptionskip}{-5pt}
    \includegraphics[width=.93\columnwidth]{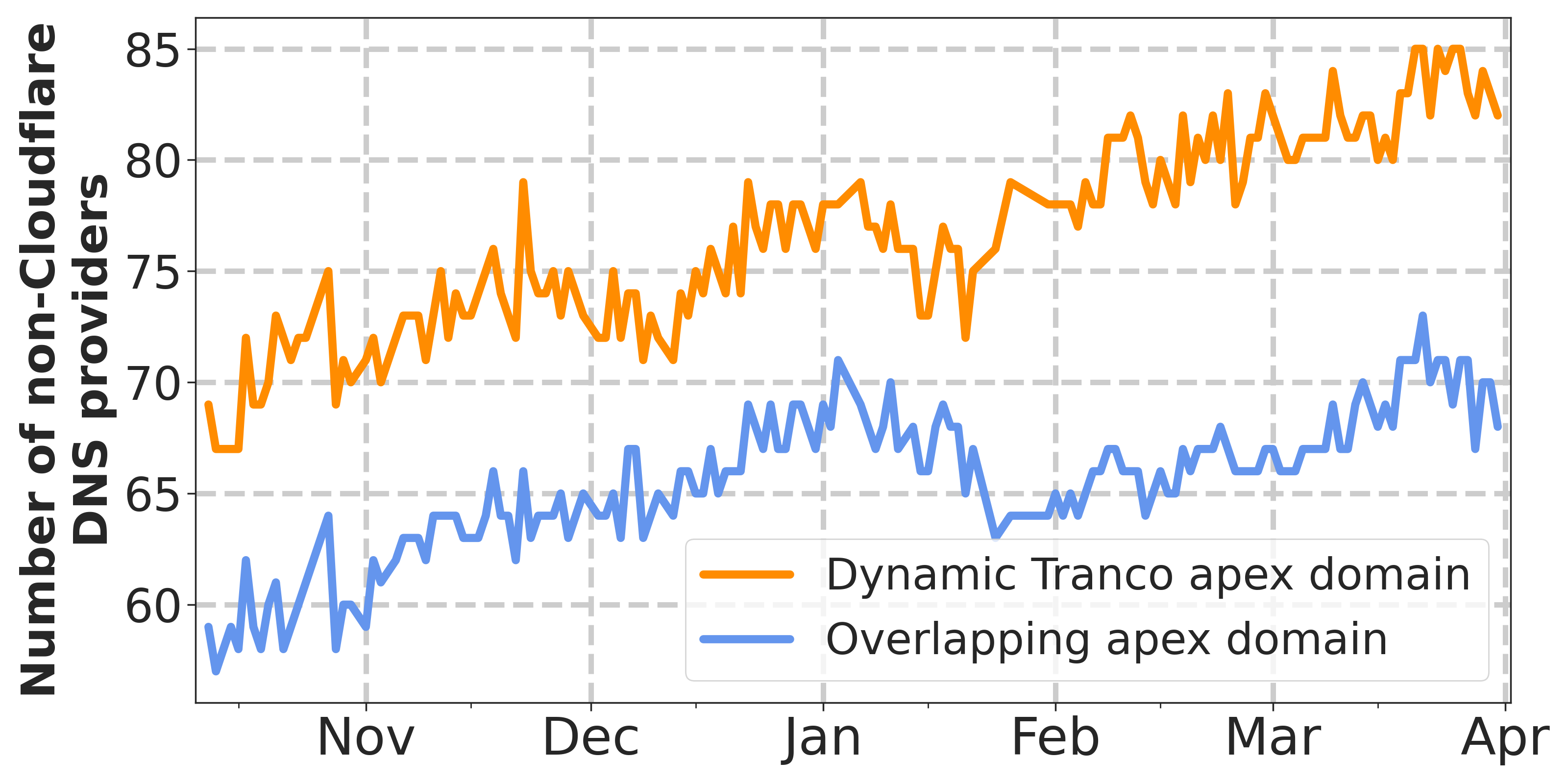}
    \caption{Number of non-\cf DNS providers employed by domains that activate \https records. }
    \label{fig:otherns-count}
\vspace{-2mm}
\end{figure}

\paragraph{\cf and non-\cf name servers.}
Interestingly, we notice that over 99\% of both dynamic Tranco and overlapping apex domains that publish \https records use \textit{Cloudflare} name servers, as shown in Table~\ref{tab:nsinfoalldom}.  
Note that we classify it as "Full \textit{Cloudflare} name server" when an apex exclusively use \textit{Cloudflare} name servers. 
On the other hand, we categorize it as "Partial \textit{Cloudflare} name server" if an apex uses a combination of \textit{Cloudflare} and non-\textit{Cloudflare} name servers. 
The "None \textit{Cloudflare} name server" indicates that an apex only utilizes name servers not provided by \textit{Cloudflare}.

In total, there are 244 distinct non-\cf DNS service provid-ers used by dynamic Tranco apex domains and 201 by overlapping apex domains that activate DNS \https records.
Figures~\ref{fig:otherns-count} illustrates an overall upward trend in the number of non-\cf DNS providers with \https records for dynamic Tranco and overlapping apex domains, respectively, ranging from 55 to 85. 
Table~\ref{tab:otherns} shows the top non-\cf DNS providers with \https records.\footnote{Akamai only supports \https records on its Edge DNS service~\cite{akamai-edgedns}, 
primarily used for hosting TLD registries. This is consistent with our finding that no \https RR is configured by Akaimai for top 1M domains.}
Further details on domains with non-\cf providers can be found in \Cref{appendix:noncf}.



\begin{table}[t]
\centering
\resizebox{0.94\columnwidth}{!}{%
\begin{tabular}{cc|cc}
\multicolumn{2}{c|}{\textbf{\begin{tabular}[c]{@{}c@{}}Dynamic\\ Tranco top 1M\end{tabular}}} & \multicolumn{2}{c}{\textbf{Overlapping}} \\
DNS provider                                      & \#. distinct domains                               & DNS provider        & \#. distinct domains        \\ \hline
eName                                      & 185                                       & GoDaddy        & 59                 \\
Google                                        & 159                                    & Google         & 40                 \\
GoDaddy                                           & 105                                & NSONE          & 20                 \\
NSONE                               & 79                                       & Hover             & 11                 \\
Domeneshop                                           & 16                      & Domeneshop              & 6                
\end{tabular}
}
\caption{Top non-\cf DNS providers 
from October 11th, 2023, to March 31st, 2024. }
\label{tab:otherns}
\vspace{-25pt}
\end{table}

\subsubsection{Inconsistent use of \https records.}
\label{subsubsec:inconsistent}

During our NS measurement period (Oct 11th, 2023 to Mar 31st, 2024), the majority of apex domains with \https records consistently maintained the \https records once published.
However, we observe 4,598 apex domains that intermittently activated \https records. 
We investigate this intermittent use of \https records by domains in relation to their name servers.
Among these 4,598 apex domains, 2,719 (59.13\%) domains consistently use the same name servers. 
Among the 2,719 domains, 2,673 (98.31\%) exclusively utilize \textit{Cloudflare} name servers, while 46 (1.69\%) employ either non-\cf name servers or both \textit{Cloudflare} and non-\textit{Cloudflare} name servers.

\paragraph{Employment of multiple name servers.}
We find that 1,593 apex domains exclusively utilize \cf name servers for activating their corresponding \https records. However, during deactivation, they utilize a mixture of \cf and other name servers. Considering that public DNS resolvers may employ specific mechanisms for selecting the most suitable DNS authoritative servers to query, we conduct additional scans by directly querying \https records from corresponding DNS authoritative servers for all apex domains. 
We then identify 6 apex domains with fewer returned \https records than corresponding name servers, due to the fact that they employ a combination of DNS providers that both support and do \emph{not} support \https records. \https records are consistently returned when querying name servers that support \https records, while none are returned by those that do not support \https records. These findings suggest that employing a mix of multiple DNS service providers, where not all provide support for \https records, may lead to inconsistent activation of \https records due to public resolvers' selecting mechanisms.

\paragraph{Change of name servers.}  
We observe that 236 apex domains lose their \https records when they switch their name servers from \textit{Cloudflare} to non-\cf name servers (172 unique ones in total). 
Additionally, 20 apex domains do not have corresponding name server records (i.e., \ns records) when they deactivate \https records, and these domains are all found to employ \textit{Cloudflare} name servers when they have \https records. 
These results suggest that one likely reason for having intermittent \https records is due to the change of name server, e.g., where the new name server does not support \https records by default.

\paragraph{Change in configuration.} 
We observe 2,673 domains who exclusively use \cf name servers despite having intermittent \https records. 
One likely reason is that \cf will automatically generate an \https record for the domain when it has the "proxied" option turned on. Thus, if the domain turns the "proxied" option on and off, it will result in intermittent \https records. 
We further investigate such default configuration by \cf next in \Cref{subsubsec:cloudflare}. 


\begin{tcolorbox}[width=\linewidth, sharp corners=all, colback=black!7, left=0.5pt, right=0.5pt, top=0.1pt, bottom=0.2pt]

\paragraph{(\textit{Takeaway})}
Though \cf remains the major adopter of \https records, there is a noticeable upward trajectory in support for \https among other prominent DNS providers.
However, using multiple DNS service providers where not all of them support \https records may result in inconsistent and intermittent \https records for a domain.

\end{tcolorbox}

\subsection{\https RR Parameters}
\label{subsec:httpsrr_param}

We now investigate how domains configure parameters in their \https records. 
Given \cf's dominating presence, 
we examine \cf name servers and non-\cf name servers separately.

\subsubsection{\textit{Cloudflare} \https RR practices}
\label{subsubsec:cloudflare}


We investigate the \https record configuration process of \cf by registering our domain to \cf's DNS service in January 2024 using a free account (features and default settings may differ with paid plans). 
\cf offers a \texttt{proxied records} feature~\cite{cloudflare-proxied-records}.
This feature, when enabled by a service user (i.e., domain owner), redirects all traffic destined for the original host to \cf's proxy server, which then forwards it to the original host. For instance, if a domain owner enables the proxied function for its \arec record, DNS queries for that record will resolve to \cf's Anycast IP addresses instead of the original IP addresses set by the domain owner.
This feature is enabled by default when a domain owner adds a new DNS record via \cf's DNS configuration dashboard.\footnote{A domain owner can explicitly disable this feature by switching off the enabled toggle.}
Moreover, once the proxied feature is activated, \textit{an \https record is automatically added} for a domain with default parameters (IP hints are specified as the IP addresses of \textit{Cloudflare}'s Anycast IPs):

\vspace{2mm}
\noindent
\fbox{\begin{minipage}{17em}
\begin{flushleft}
\small\texttt{a.com. 300 IN HTTPS 1 . alpn=h2,h3 ipv4hint=a.b.c.d ipv6hint=e:f::g}
\end{flushleft}
\end{minipage}}
\vspace{2mm}

\noindent Furthermore, this default \https record cannot be modified and other \https records cannot be added if a domain is using the proxied feature.
A domain can only set its own \https records after disabling the proxied feature.

Based on this information, we divide domains employing \cf name servers into two groups: those with default \cf \https RR configuration and those with customized \https RR configuration. 
We identify each domain's category by comparing its \https record configuration against \textit{Cloudflare}'s default configuration for free users.
The corresponding ratios are presented in \Cref{tab:cfconfigratio}.

\begin{table}[!htbp]
\vspace{-1mm}
\small
\begin{tabular}{c|c|c}
\textbf{}     
\textbf{\begin{tabular}[c]{@{}c@{}}\https RR\\Configuration\end{tabular}} & \textbf{\begin{tabular}[c]{@{}c@{}}Dynamic\\ Tranco top 1M (\%)\end{tabular}} & \textbf{Overlapping (\%)} \\ \hline
\textbf{Default}   & 79.96    & 72.37    \\ \hline
\textbf{Customized} & 20.04    & 27.63    \\                  
\end{tabular}
\caption{The average ratio of domains using \cf name servers that have default \https record configuration v.s. customized configuration.}
\label{tab:cfconfigratio}
\vspace{-6mm}
\end{table}

We observe that over 79\% and 72\% of dynamic and overlapping domains, respectively, adhere to the default \https record configuration provided by \textit{Cloudflare}.
On the other hand, 
the remaining 20\% and 28\% of dynamic and overlapping domains, which have customized \https configuration parameters, may likely be aware of their \https records. 
Thus, this portion might serve as a conservative estimate for the number of domains potentially aware of their \https record usage. 
Hence, in our following analysis, we distinguish between domains that utilize \textit{Cloudflare}'s default \https record configuration and those with customized configuration.

\subsubsection{Other DNS providers' \https RR practices}
\label{subsubsec:otherns}

We investigate the configuration of \https records associated with name servers from \textit{Google} and \textit{GoDaddy} as well, considering their popularity among DNS providers utilized by domains that support \https records (See Table~\ref{tab:otherns}). 
We show \https record configurations and corresponding ratios in Table~\ref{tab:othernsconfig}.

\begin{table}[!htbp]
\resizebox{0.96\columnwidth}{!}{%
    \begin{tabular}{c|c|c}
    \textbf{}     
      & \textbf{\textit{Google} NS} & \textbf{\textit{GoDaddy} NS} \\ \hline
    \textbf{\sprty}   & 1 (98.95\%)  & 0 (99.19\%) \\ \hline
    \textbf{\tname} & . (98.95\%)  & An alternative endpoint (99.19\%) \\
    \hline
    \textbf{\alpn} & - (95.11\%) & - (99.19\%) \\
    \hline
    \textbf{\ipfh} & - (97.76\%) & - (99.19\%) \\
    \hline
    \textbf{\ipsh} & - (98.90\%) & - (99.19\%) \\
    \end{tabular}
    }
\caption{Common \https configurations by \textit{Google} and \textit{GoDaddy} name servers. Percentages indicate the ratio of domains from each name server with the specified configuration parameter. Dashes denote empty fields.}
\label{tab:othernsconfig}
\vspace{-8mm}
\end{table}

While the majority of \https records with \textit{Google} name servers are in \smode, most of them have empty \sparam fields, therefore not offering additional domain information. 
Conversely, we observe that 558 \https records specify \alpn, predominantly supporting HTTP/2, and 172 records incorporate \ipfh. 
Note that one domain may be associated with multiple \https records.
Among all these domains, aside from 153 that are owned by \textit{Google}, we observe 10 not affiliated with \textit{Google}.\footnote{\texttt{cromwell-intl.com}, \texttt{fetlife.com}, \texttt{err.ee}, \texttt{wakeuplaughing.com}, \texttt{pixelcrux.com}, \texttt{stvincenttimes.com }, \texttt{dukarahisi.com}, \texttt{miranajewels.com}, \texttt{americancensorship.org}, and \texttt{smalls.com}.}
Notably, \texttt{err.ee} is the sole apex domain that utilizes \amode, aliasing to its \www subdomain.

In contrast, most \https records utilizing \textit{GoDaddy} name servers are configured in \amode, redirecting to alternative endpoints. 
Among the remainder, which amount to 44 apex domains, 36 support both HTTP/3 and HTTP/2, while 8 exclusively support HTTP/2. 
Furthermore, 42 out of these 44 domains incorporate both \ipfh and \ipsh in their \https records. 


\subsubsection{SvcPriority and TargetName}
\label{subsubsec:param_target}

As explained in \Cref{sec:background}, an \https records can be deployed in two distinct modes: \amode (value 0) which aliases a domain to the other domains, and \smode (other values) which provides information associated with a specific service endpoint.

We observe that the vast majority of domains use \sprty value of 1 (i.e., \smode) with both \cf name servers (99.97\% and 99.95\% of overlapping apex and www domains) and non-\cf name servers (96.65\% of overlapping apex domains). 
However, 202 apex domains do not provide any \sparam even though they are in \smode. On the other hand, among the remaining domains in \amode, we find that 19 domains set themselves as the \tname (i.e., by using ``.'' as value), which does not appropriately provide a true alias.
Further details in \Cref{appendix:param_target}.




\subsubsection{ALPN}
\label{subsubsec:alpn}

A domain can specify the application protocols it supports through the \alpn parameters in its \https records.
We observe that the prevalent protocols in \alpn are HTTP/2 and HTTP/3, which is attributed to \cf's default configuration (\Cref{subsubsec:ns}). 
For domains with non-\cf name servers, we observe a much lower ratio of advertising HTTP/2 and HTTP/3, with an average of 64.09\% and 26.79\%, respectively. 
Furthermore, 8.44\% of domains do not include \alpn parameters, and 1 domain continuously advertise draft versions 27 and 29 of HTTP/3. 
Further details are in \Cref{appendix:alpn}.

\subsubsection{IP Hints}
\label{subsubsec:iphint}
A domain can include \ipfh/\ipsh parameters (i.e., IP hints) in its \https record to specify the IP addresses that clients can use to access the service it provides. 
We observe that over 97\% and 87\% of both apex and www overlapping domains adopt \ipfh and \ipsh, respectively.
We further examine the consistency between the IP addresses provided in the IP hints and those in the corresponding \arec/\aaaa records of the domains. 
We find that 624 and 5,109 of apex and www domains, respectively, have exhibited inconsistency, with an average of 6.57 and 14.52 days in duration of the inconsistency. Further details can be found in~\Cref{appendix:iphint}.

\paragraph{Connectivity of domains with mismatched IP hints.}
One question that naturally arises when there is a mismatch between IP hints and \arec/\aaaa is whether connections can be successfully established to both IPs. 
To answer this question, we conduct an additional experiment from January 24th to March 31st, 2024.
For any apex domain with \https records, we perform another sequence of queries of its \https, \arec and \aaaa records. After receiving the DNS responses, we promptly examine the consistency of its IP addresses and initiate TLS handshakes with all IP addresses in the \https, \arec and \aaaa responses through an OpenSSL client, if the domain has mismatched IP hints. This allows for an immediate validation of connectivity.

We identify a total of 1,022 occurrences of domains exhibiting IP mismatches (a domain is counted multiple times if it displays mismatched IP addresses on different days), of which 317 are distinct domain names.
5 domains\footnote{\texttt{cf-ns.net}, \texttt{cf-ns.com}, \texttt{canva-apps.cn}, \texttt{cloudflare-cn.com}, \texttt{polestar.cn}} consistently demonstrate IP mismatches throughout the entire observation period. Subsequent TLS handshake attempts to these domains after each occurrence reveal that 193 domains have at least one unreachable IP address in either their IP hints and/or \arec/\aaaa records, most commonly with an unreachable network error. 
Among these 193 domains, 117 can only be reached through the IP addresses in their IP hints, whereas 59 domains are accessible solely via their \arec record.

Our observations regarding the inconsistencies between IP hints and \arec/\aaaa records highlight the complexities involved in managing \https records.
Domains need to synchronously update both their IP hints (in \https records) and \arec/\aaaa records whenever they change their IP addresses.
However, given that most inconsistencies are resolved within a few days, such mismatches may often stem from \textit{DNS caching effects}.
For instance, even though a domain simultaneously updates both its \https and \arec records, recursive DNS resolvers may continue to serve cached records until their TTL (Time to Live) expires.
Therefore, inconsistency issues can emerge when the expiration timings of these cached records differ.

Moreover, our connectivity experiments indicate that these inconsistencies can make domains unreachable for clients if clients fail to utilize both \https records and \arec/\aaaa records. 
In \Cref{sec:client}, we investigate the browsers' behaviors in response to these issues.

\begin{tcolorbox}[width=\linewidth, sharp corners=all, colback=black!7, left=0.5pt, right=0.5pt, top=0.1pt, bottom=0.2pt]

\paragraph{(\textit{Takeaway})}
The use of IP hints (\ipfh/\ipsh) adds an additional layer of complexity in maintaining IP addresses in both \https and \arec/\aaaa records. Incorrect IPs could make domains unreachable for clients, especially if clients do not implement robust failover mechanisms. 

\end{tcolorbox}

\input{sections/4_server_dnssec_ech}

%% file: tables/tab-dataset.tex
\begin{table*}[]
\footnotesize
\begin{tabular}{cc|c|cccccc}
\multicolumn{2}{c|}{\multirow{3}{*}{\textbf{Data Type}}} &
  \multirow{3}{*}{\textbf{Measurement Period}} &
  \multicolumn{6}{c}{\textbf{Utilized in Section}} \\
\multicolumn{2}{c|}{} &
   &
  4.2.1 &
  4.2.2 &
  4.2.3 &
  4.3 &
  4.4 &
  4.5 \\
\multicolumn{2}{c|}{} &
   &
  \begin{tabular}[c]{@{}c@{}}Adoption \\ rates\end{tabular} &
  \begin{tabular}[c]{@{}c@{}}Name \\ servers\end{tabular} &
  \begin{tabular}[c]{@{}c@{}}Inconsistent \\ use\end{tabular} &
  \begin{tabular}[c]{@{}c@{}}\https RR \\ parameters\end{tabular} &
  \begin{tabular}[c]{@{}c@{}}ECH \\ deployment\end{tabular} &
  \begin{tabular}[c]{@{}c@{}}\https RR \\and DNSSEC\end{tabular} \\ \hline
\multicolumn{1}{c|}{\multirow{2}{*}{\textbf{\begin{tabular}[c]{@{}c@{}}Domain\\ (Apex, \www)\end{tabular}}}} &
  \https, \arec, \aaaa &
  2023-05-08 -- 2024-03-31 &
  \checkmark &
  \checkmark &
  \checkmark &
  \checkmark &
  \checkmark &
  \checkmark \\ \cline{2-9} 
\multicolumn{1}{c|}{} &
  \soa, \ns &
  2023-08-16 -- 2024-03-31 &
   &
  \checkmark &
  \checkmark &
  \checkmark &
   &
   \\ \hline
\multicolumn{1}{c|}{\textbf{Name Server}} &
  \arec, \aaaa, WHOIS &
  2023-10-11 -- 2024-03-31 &
   &
  \checkmark &
  \checkmark &
   &
   &
   \\ \hline
\end{tabular}
\caption{Overview of datasets for server-side analysis. } 
\label{tab:server_dataset}
\vspace{-6mm}
\end{table*}

%% file: sections/4_server_dnssec_ech.tex

\subsection{ECH Deployment}
\label{subsec:ech}
An important feature of \https records is their ability to deliver TLS Encrypted Client Hello (ECH) configurations, 
which enables clients to send encrypted \chello messages to the server (associated with the domain).

\subsubsection{ECH support }
\label{subsubsec:echsupport}
We identify domains with ECH parameter specified in their \https records, and uncover a sudden change in ECH support on October 5th, 2023. 

\paragraph{Before October 5th, 2023.}
The overall trend remains relatively stable, with about 70\% of apex domains and 63\% of \www subdomains utilizing \https records having adopted ECH. 
Considering that \cf automatically activates ECH for free zones prior to October 5th, 2023~\cite{cloudflare-ech},
the substantial deployment of ECH for both apex domains and \www subdomains is consistent with our earlier observations that a significant portion of domains using \https RR employ \cf name servers with the \texttt{proxied} option enabled (\Cref{subsubsec:cloudflare}).
Specifically, 99.95\% of ECH-enabled apex domains and \www subdomains use \cf name servers.

ECH was also activated by 106 apex domains and 74 \www subdomains in conjunction with 48 non-\cf name servers.\footnote{The top three most utilized non-\cf name servers are \texttt{ubmdns.com}, \texttt{domainactive.org}, and \texttt{informadns.com}.}
Interestingly, we find that \textit{all} ECH configurations used by these domains direct to the same \cf server, regardless of their name servers' DNS providers. 
We discuss this in further detail in Section~\ref{subsubsec:echdns}.




\paragraph{After October 5th, 2023.}
A notable drop in the number of domains with ECH is observed on October 5th, 2023, resulting in \textit{zero domains with ECH}.\footnote{Cloudflare continues to operate one or two domains for testing ECH beyond October 5th, 2023. We exclude these domains, \texttt{cloudflare-ech.com} and \texttt{cloudflareresearch.com}, from our daily counts.}
Our finding was confirmed by Cloudflare's announcement that ECH features were disabled for all domains using their services~\cite{cloudflare-ech-down}, due to "a number of issues".\footnote{While Cloudflare's announcement mentioned ECH re-enablement in early 2024, it has not been re-enabled at the time of this writing.}
While we cannot directly confirm the exact reasons that lead to \cf's sudden disabling of ECH, we investigate potential challenges and issues in ECH usage---from both server-side (\Cref{subsubsec:echdns}) and client-side (\Cref{subsec:browser_ech})---that can shed light on the future of ECH deployment.  

\subsubsection{Managing ECH with DNS caches}
\label{subsubsec:echdns}

Since ECH configurations are delivered via \https records and these records are cached by recursive DNS resolvers (or stub resolvers), domains have to carefully maintain their ECH configurations (i.e., ECH keys), taking into account \textit{DNS caches (i.e., cached \https records)}.
When the server changes its ECH configuration, it should maintain both the \https record with the previous ECH configuration and the one with the new configuration to account for resolvers that may still have the old configuration in the cache.
Otherwise, it could lead to inconsistencies between the ECH key delivered to a client via (cached) \https records and the actual key recognized by the server.
Similar types of key rotation (or key rollover) are known to be challenging~\cite{chung2017longitudinal,lee2022under}.

\begin{figure}[t!]
    \centering
    \setlength{\abovecaptionskip}{2pt}
    \setlength{\belowcaptionskip}{-5pt}
    \includegraphics[width=.95\columnwidth]{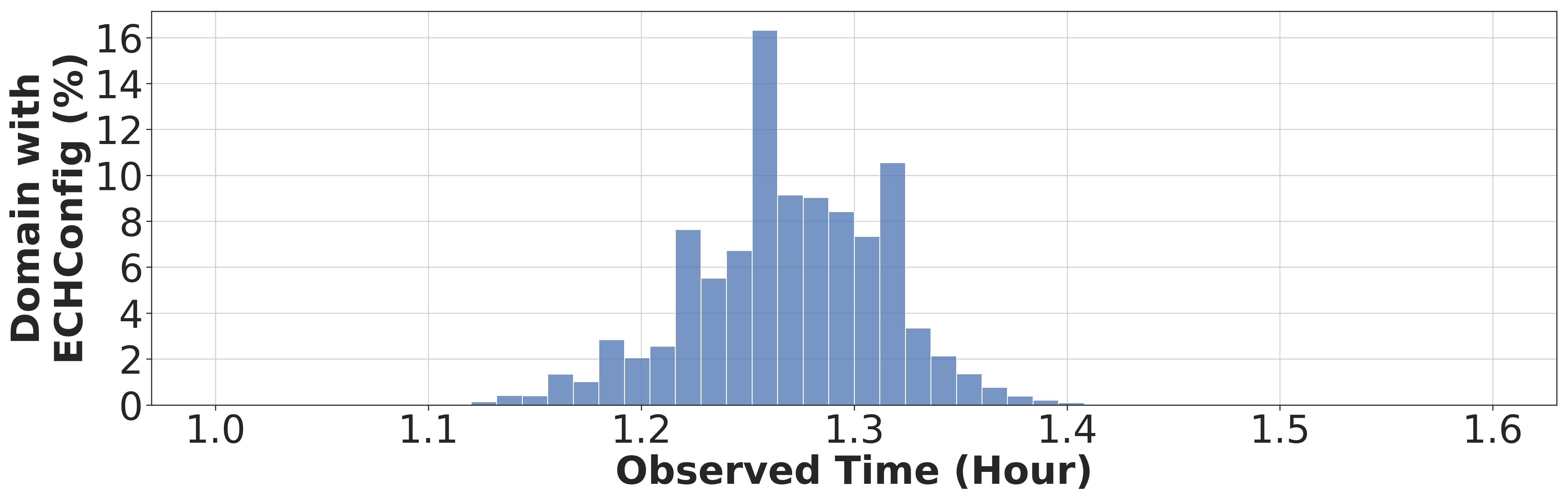}
    \caption{
    Percentage of domains based on the average duration of their ECH configuration.} 
    \label{fig:ech-rotation}
\vspace{-3mm}
\end{figure}

To further investigate the ECH key rotation frequency in \https records, we conduct hourly scans on Tranco apex domains over seven days, from July 21 to July 27, 2023. 
As a result, we identify 169 unique ECH configurations, all directing to the same public client-facing server (\texttt{cloudflare-ech.com}), each associated with a distinct public key.
Among these configurations, the majority (154) are consistently observed in two consecutive hourly scans; additionally, 13 configurations are presented in three consecutive hours and 2 are observed in only one hour.
We notice that the TTL for over 99\% of these \https records is set to 300 seconds.
\Cref{fig:ech-rotation} depicts the distribution of domains based on the average duration their ECH configuration is observed.
The duration periods range from 1.1 to 1.4 hours, with an overall average of 1.26 hours across all domains.
This observation indicates that ECH keys (in these configurations) are rotated every one to two hours.

Such key rotation frequency implies the possibility of key inconsistency that may occur approximately every one to two hours,
given the complexity of managing ECH keys in conjunction with DNS caches.
As elaborated in \Cref{subsubsec:iphint}, IP hints parameters (\ipfh/\ipsh) in \https records face similar challenges with DNS caches.
Therefore, to properly make use of ECH, it requires both servers to implement retry configuration and clients to correctly handle it (detailed in \Cref{subsec:browser_ech}).
Otherwise, it may prohibit encrypted TLS \chello messages between servers and clients, and even the failure of the connection.


\begin{tcolorbox}[width=\linewidth, sharp corners=all, colback=black!7, left=0.5pt, right=0.5pt, top=0.1pt, bottom=0.2pt]

\paragraph{(\textit{Takeaway})} 
Managing ECH unavoidably adds additional complexity to the servers, especially considering the frequent key rotation. Both servers and clients need to handle ECH keys properly to avoid connection failures. 

\end{tcolorbox}


\subsection{\https RR and DNSSEC}
\label{subsec:dnssec}







DNSSEC~\cite{rfc4033,rfc4034,rfc4035} ensures the integrity and authenticity of DNS records.
DNSSEC introduces three DNS record types; \dnskey, \rrsig (Resource Record Signature), and \ds (Delegation Signer) records.
Although DNSSEC is optional in the current \https record specification, 
it is critical to deploy DNSSEC for \https records. 
Otherwise, it is susceptible to being dropped or forged by attackers, posing similar security risks to those encountered in traditional HTTP to HTTPS redirection (due to their plaintext transmission).

\subsubsection{Signed \https records}
\label{subsubsec:dnssec_deployment}

\begin{figure}[ht]
\vspace{-3mm}
    \centering
    \setlength{\abovecaptionskip}{5pt}
    \setlength{\belowcaptionskip}{1pt}
    \begin{subfigure}[b]{.99\columnwidth}
        \includegraphics[width=.99\columnwidth]{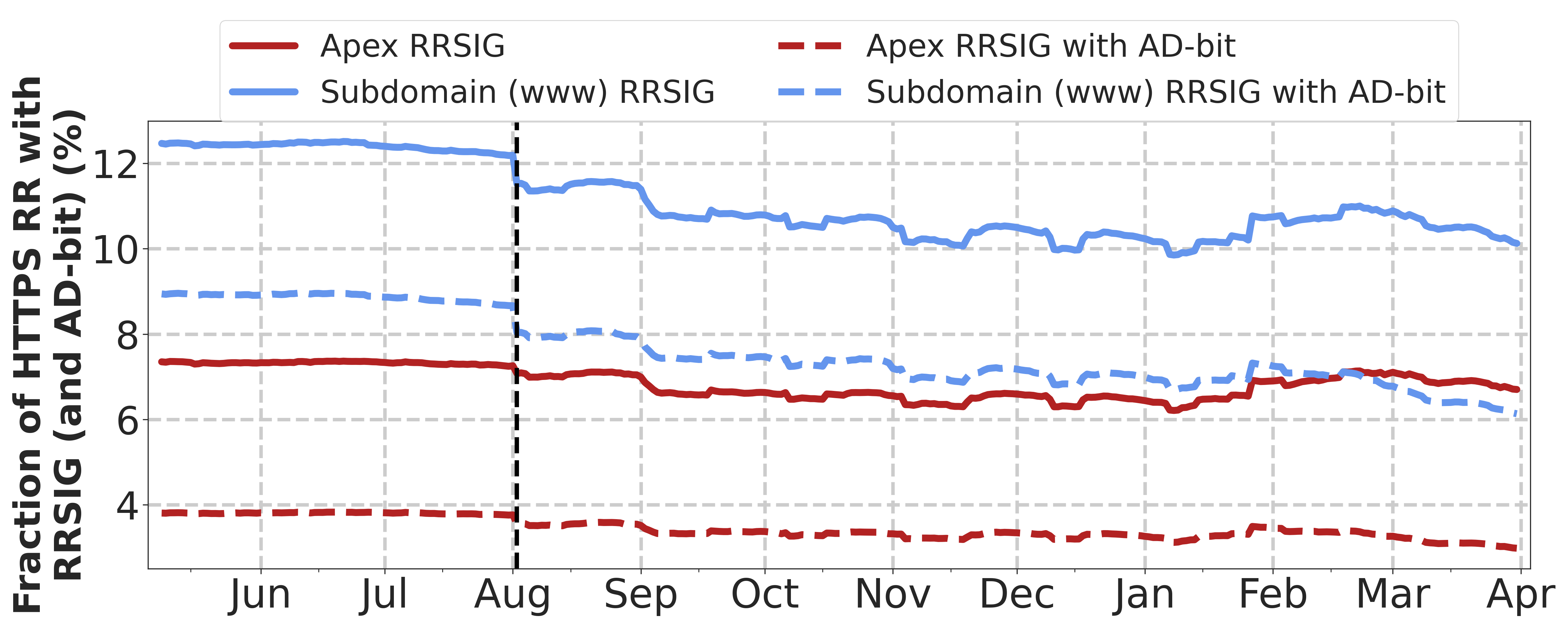}
        \caption{The fraction for dynamic Tranco top 1M domains.}
        \label{fig:adbit-all}
    \end{subfigure}
    \vspace{-2mm}
    \begin{subfigure}[b]{.99\columnwidth}
        \includegraphics[width=.99\columnwidth]{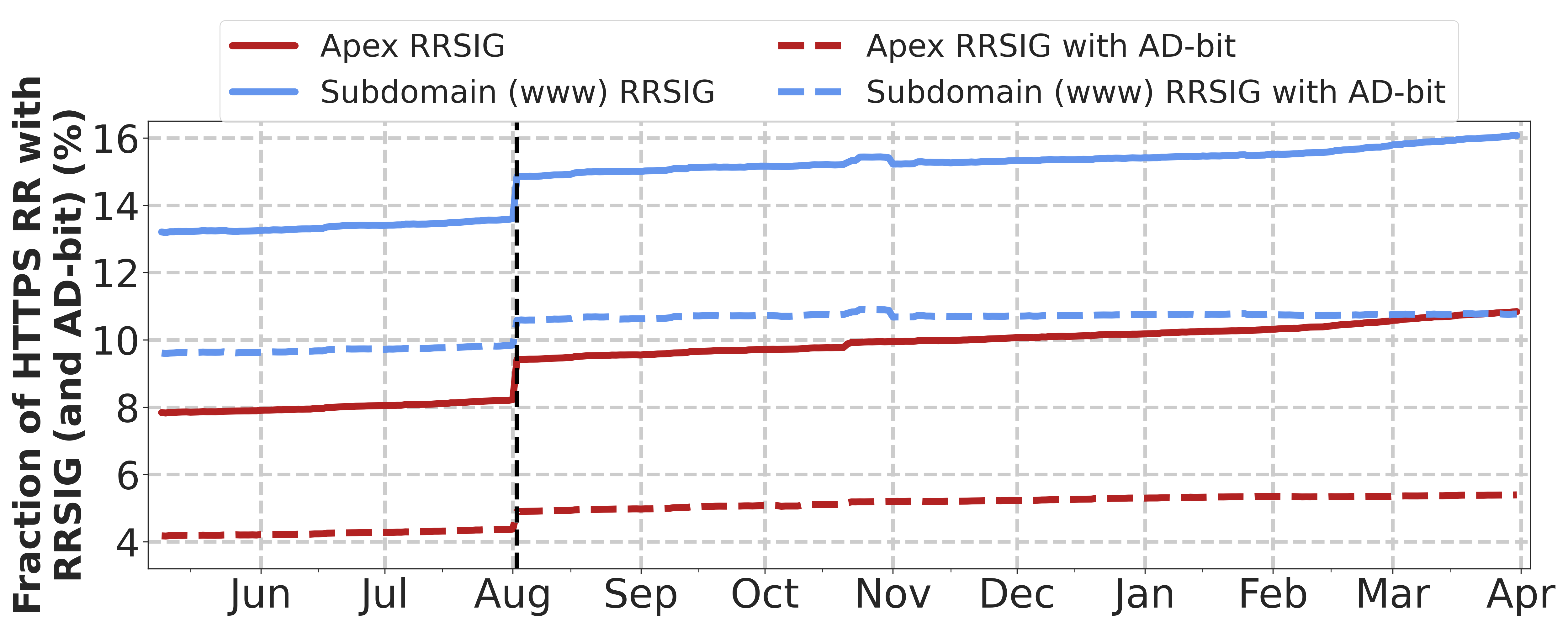}
        \caption{The fraction for overlapping domains.}
        \label{fig:adbit-overlap}
    \end{subfigure}
    \caption{Percentages of \https records with \rrsig (solid line), \rrsig and \adb bit (dashed line). Vertical dashed line (on August 1st, 2023) denotes Tranco source change.}
    \label{fig:rrsig}
\vspace{-3mm}
\end{figure}

In our daily scan of \https records, we simultaneously collect the corresponding \rrsig records, if provided by the servers.
Using the collected \rrsig records, we examine the DNSSEC support of domains with \https records.
\Cref{fig:rrsig} depicts the ratio of \emph{signed} \https records (i.e., \https records that have the corresponding \rrsig records) in solid lines, and the ratio of \emph{validated} \https records (i.e., the \texttt{Authenticated Data (AD)} bit~\cite{rfc3655} is set in the DNS response) in dashed lines. 
We note that the observed ratio of \rrsig aligns with the trends in DNSSEC deployment~\cite{yajima2021measuring}. 

\paragraph{Prevalence of signed \https RR.}
The ratio of signed \https records shows a decreasing trend for both apex and \www subdomains (solid lines in \Cref{fig:adbit-all}) when using dynamic domain list.
In contrast, the overlapping domains show increasing trends (solid lines in \Cref{fig:adbit-overlap}), 
indicating that the deployment of DNSSEC for \https RR is growing for domains that are consistently in Tranco 1M. On the other hand, the relatively lower-ranked and possibly newly-added domains in the dynamic domain list are less likely to have DNSSEC deployed already, contributing to the overall decreasing trend. 

Interestingly, this trend is reversed for the overall \https records deployment (in~\Cref{fig:httpsrr-adopt}); there is an increasing trend in \https records deployment for dynamic domains but a decreasing trend for overlapping domains.
This suggests that relatively higher-ranked domains are more likely to support DNSSEC than \https records, possibly due to the recency of \https records. 
\paragraph{Validity of signed \https RR.}
We examine the validity of the \rrsig records of \https records by utilizing the \adb bit~\cite{rfc3655} in the DNS responses for \https records. The \texttt{AD} bit is set by a recursive resolver only if the DNSSEC chain of the corresponding DNS record has been verified.
\Cref{fig:rrsig} illustrates the ratio of validated \https records in dashed lines, 
which is much lower than that of the signed records. 
For example, for overlapping apex domains, 47.8\% of signed \https records cannot be validated likely due to issues in their DNSSEC chain, leaving the authenticity of these \https records in question.
Upon further analysis, we find that the lack of validation is likely due to the well-known issue where domains use a third-party DNS operator instead of their registrar's DNS service, and consequently fail to upload necessary \ds records themselves~\cite{chung2017understanding}. 
In our dataset, only 26\% of apex domains with signed \https records use the same DNS operator and registrar.
Further details are in \Cref{appendix:dnssec-apex}.



\subsubsection{ECH with DNSSEC} 

The low adoption rate of DNSSEC among domains publishing \https records could additionally pose security concerns to the use of ECH. 
It is ironic that ECH delivers the server's public key to the client, and yet in the absence of DNSSEC, such key cannot be fully trusted.
%
%
%
Unfortunately, our data shows that the vast majority of \https records with ECH parameters are not signed, 
Before October 5th, 2023 (the date \cf disabled ECH), less than 6\% of overlapping domains with \https and ECH are signed, and only half of them can be validated.


\begin{tcolorbox}[width=\linewidth, sharp corners=all, colback=black!7, left=0.5pt, right=0.5pt, top=0.1pt, bottom=0.2pt]

\paragraph{(\textit{Takeaway})} 
We observe very limited (<10\%) DNSSEC deployment for \https records, 
nearly half of which 
cannot be validated due to missing \ds records. The lack of proper DNSSEC deployment leaves the majority of \https records susceptible to attacks such as DNS cache poisoning. 

\end{tcolorbox}

%% file: sections/5_client.tex
\section{Client-side \https RR Support}
\label{sec:client}





So far, we have observed that over 20\% of the top 1 million domains have adopted \https records, with the majority employing various parameters, such as \sprty, \alpn and IP hints. 
However, the effectiveness of these records 
depends significantly on their proper utilization by clients.
Thus, an important question arises: how do clients, particularly web browsers, support the \https records? In this section, we aim to examine the support of \https records across popular web browsers~\cite{browsermarket}, including Chrome, Safari, Edge,\footnote{Both Edge and Chrome are built upon the Chromium~\cite{chromium} framework. However, we perform separate experiments on these browsers, as variations in their custom implementations may result in differing behaviors. } and Firefox, to evaluate the potential impact on domains that have adopted these records. 
Specifically, we investigate (1) whether the browsers support \https record lookup, (2) whether the browsers utilize the fetched \https records in establishing connections, and (3) how they respond to misconfigured \https records (i.e., failover behavior).
Furthermore, we perform an in-depth analysis to assess the ECH support from the popular browsers.



\begin{figure}[!t]
    \centering
    \setlength{\abovecaptionskip}{2pt}
    \setlength{\belowcaptionskip}{-5pt}
    \includegraphics[width=0.95\columnwidth]{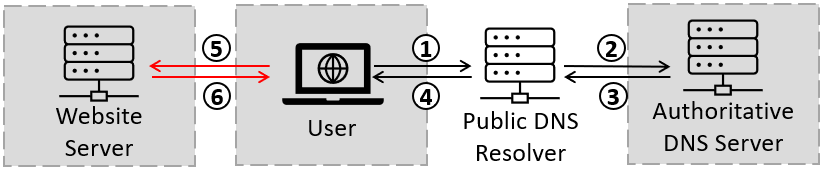}
    \caption{Experimental setup for client-side browser behavior analysis. Shaded regions denote controlled environments, namely the testing client, authoritative DNS server, and the target domain website server. }
    \label{fig:client-exp}
\vspace{-2mm}
\end{figure}

\paragraph{Experimental setups.}
To test web browsers' \https record support, we set up a testbed, as illustrated in \Cref{fig:client-exp}.
We configure our own domain and operate an authoritative name server using BIND9~\cite{bind9} to configure various \https record settings for our domain.
Additionally, we set up a web server (accessible through our domain name) that supports ECH, utilizing the publicly available ECH (draft version 13) implementation of OpenSSL~\cite{opensslech} and Nginx~\cite{nginxech}. 
Both the name server and web server are hosted on Amazon Web Services (AWS) instances.
Target web browsers run with default settings on separate machines with Microsoft Windows 11 (Home edition) and macOS Sonoma; Safari tests are omitted on Windows due to its lack of support. 
For DNS resolution, we utilize Google DNS resolver. For Firefox, we set its DoH server as: https://dns.google/dns-query.
We choose to use a public resolver
due to its widespread use and responsiveness, 
allowing us to focus on evaluating the browsers' \https record support without the added complexity of varying resolver behaviors. 
Note that our test environment is configured for IPv4 addresses, and the interaction of browsers with IPv6 addresses remains as future work. 

To eliminate the cache effect, in each round of the experiment, we begin by clearing the local DNS cache and resetting the browser history. 
We set the DNS record's Time-to-Live (TTL) to 60 seconds to ensure frequent refreshes of DNS records by the public resolver. Between experiments, we ensured that the interval exceeded the 60-second TTL to prompt public resolvers to fetch fresh DNS records (e.g., \https records) from our authoritative DNS server. 
The test flow is as follows, as illustrated in~\Cref{fig:client-exp}.
Upon instructing the browser to access our domain, the browser sends DNS queries (e.g., for \arec or \https record) to the DNS resolver (\ding{172}), which then sends queries (\ding{173}) to our authoritative name server (associated with our domain).
Subsequently, the corresponding DNS response will be returned to our test browser (\ding{174}$\sim$\ding{175}).
Finally, the browser starts the TLS handshake process with our web server (\ding{176}$\sim$\ding{177}).
We repeat this process 5 times for each target browser and for each parameter setting.

\input{tables/client-behavior}

\subsection{Support by Popular Browsers}
We first measure the overall support for \https RR across four web browsers. 
%
%
One objective of \https RR is to signal that the HTTPS protocol should be employed instead of the HTTP protocol
when a client (e.g., browser) connects to a host (e.g., web server). 
To evaluate browser support for \https record, we directed browsers to access three types of URLs: (i) \texttt{a.com}, (ii) \texttt{http://a.com} and (iii) \texttt{https://a.com}. 
The purpose of the first two URLs is to assess the utilization of the \https record.
For instance, following the retrieval of the \https record, if a browser directly initiates an HTTPS connection when accessing the first two URLs, it indicates that the browser utilizes the \https record as a signal of the web server's support for HTTPS.
Therefore, this experiment aims to verify (1) whether browsers issue DNS queries for \https RR and (2) whether they leverage the response to initiate an HTTPS connection as per the standard~\cite{rfc9460}.

Our test domain has the following \https record: 
\smode (\sprty set to 1), \tname pointing to itself (value ``.''), and \alpn specifying support of \texttt{h2} (HTTP/2). 
We also publish the \arec record that points to our web server.

\vspace{1.5mm}
\noindent
\fbox{\begin{minipage}{15em}
\begin{flushleft}
\small\texttt{a.com. 60 IN HTTPS 1 . alpn=h2 }\\
\small\texttt{a.com. 60 IN A 1.2.3.4}
\end{flushleft}
\end{minipage}}
\vspace{1.5mm}

The results are summarized in~\Cref{tab:client-https}, inside "\https RR Utilization".
First, we observe that all the tested browsers request two distinct DNS queries, i.e., for \https records and for \arec records, when visiting all three types of URLs.
An \https RR query is issued even when the target server lacks a corresponding \https record, as the browser is unable to ascertain this beforehand. 
However, only Chrome, Edge, and Firefox\footnote{Firefox only queries HTTPS records if using DNS over HTTPS (DoH)~\cite{firefox-doh}, which is enabled by default, while Chrome does not require DoH for HTTPS. Note that DoH is needed to  preserve the privacy of the domain name (in addition to ECH), and DNSSEC is still needed to provide full protection. }
proceed to initiate HTTPS connection with web server across all URL variations, indicating their actual use of the \https records as a signal for HTTPS support of the server.
In contrast, Safari still establishes HTTP connections (i.e., via port 80) for the first two URL types (\texttt{a.com} and \texttt{http://a.com}), implying that it does not utilize the fetched \https records (half circle in \Cref{tab:client-https} indicating fetching \https RR but not utilizing it).



\begin{tcolorbox}[width=\linewidth, sharp corners=all, colback=black!7, left=0.5pt, right=0.5pt, top=0.1pt, bottom=0.2pt]

\paragraph{(\textit{Takeaway})} 
While all four browsers send queries for \https records, one browser, Safari, does not utilize the fetched \https records to initiate secure connections.

\end{tcolorbox}

\subsection{Resolution of \https RR Parameters}
We next investigate how browsers utilize the parameters specified in \https records.
The \https record specification~\cite{rfc9460} states that the resolution of \https records relies on the client rather than the recursive resolver, thus we do not consider interventions from resolvers.
Given that Safari only utilizes \https records to initiate HTTPS connection when visiting \texttt{https://a.com}, we examine the behavior of all browsers by visiting our test domains with the \texttt{https://} prefix.

\subsubsection{AliasMode}
\amode (\sprty value of zero) serves a crucial purpose by enabling aliasing at the zone apex, which cannot be conducted with \cname records.
In this mode, a domain specifies another domain it wishes to point to in the \tname field.
Consequently, with an \amode \https record, a browser is expected to access the domain specified in the \tname field (i.e., by issuing \arec record queries).
To verify if browsers adhere to this anticipated behavior, we configure our DNS zone as follows:

\vspace{1.5mm}
\noindent
\fbox{\begin{minipage}{15em}
\begin{flushleft}
\small\texttt{a.com. 60 IN HTTPS 0 pool.a.com.}\\
\small\texttt{pool.a.com. 60 IN A 1.2.3.4}
\end{flushleft}
\end{minipage}}
\vspace{1.5mm}

\noindent We observe that only Safari issues subsequent DNS queries for or establishes a connection with the server (at \texttt{1.2.3.4}) that hosts \texttt{pool.a.com.}, as indicated in \Cref{tab:client-https} (\amode section).
In contrast, the other three browsers simply try to access \texttt{a.com.} but fail due to the absence of an associated IP address with \texttt{a.com}.

\subsubsection{ServiceMode}

A domain can provide information about its alternative service endpoint accessible to browsers through \smode.
Specifically, an \https record configured with \smode can include parameters such as \port, IP hints, and \alpn.
Here, we examine how browsers utilize these parameters provided in an \https record.
The \ech parameter, another critical feature of the \https record, will be discussed in the subsequent \Cref{subsec:browser_ech}.
\Cref{tab:client-https} (\smode section) summarizes our findings.

\paragraph{\tname.} 
In \smode, the \tname field specifies the domain name of the alternative service endpoints. 
We first examine if browsers utilize the domain name specified in this \tname field.
In the following setup, the \https-RR-aware browser is expected to establish an HTTPS connection with the server (\texttt{at 2.2.3.4}) which hosts \texttt{pool.a.com} using the provided information (i.e., \alpn).

\vspace{1.5mm}
\noindent
\fbox{\begin{minipage}{19em}
\begin{flushleft}
\small\texttt{a.com. 60 IN HTTPS 1 pool.a.com. alpn=h2}\\
\small\texttt{a.com. 60 IN A 1.2.3.4} \\
\small\texttt{pool.a.com. 60 IN A 2.2.3.4}
\end{flushleft}
\end{minipage}}
\vspace{1.5mm}


\noindent We notice that both Safari and Firefox adhere to the domain name in the \tname by issuing additional queries or accessing the server that hosts the domain.
However, 
Chrome and Edge fail to utilize \tname, resulting in failure of obtaining the right service.
Therefore, in the subsequent experiments, we explicitly specify the owner name (i.e., the original domain that publishes \https records) as the \tname by setting its value to ``.''.

\vspace{-1mm}
\paragraph{(1) \port.} 
Browsers are anticipated to connect to an alternative endpoint (specified in \tname) using the port included in the \port field.
The following configuration directs browsers to use port 8443 for the \https connection to the target server.
Incorrect handling or ignoring of this parameter could lead to failure of establishing the connection.

\vspace{1.5mm}
\noindent
\fbox{\begin{minipage}{19em}
\begin{flushleft}
\small\texttt{a.com. 60 IN HTTPS 1 . alpn=h2 port=8443}\\
\end{flushleft}
\end{minipage}}
\vspace{1.5mm}

\noindent Our observation reveals that neither Chrome nor Edge utilizes the \port parameter; both browsers initiate connections directly to port 443, the default port number for HTTPS services, leading to a failure to access the web service. 
In contrast, Safari and Edge successfully initiate connections using the designated port (i.e., 8443).

\paragraph{\port failover.} 
To examine browser behavior in response to connection failures, 
we set up three server configurations: one accessible only on port 443, another only on port 8443, and a third on both ports.
Chrome and Edge experience a hard failure when they cannot establish a connection on their initially attempted port (port 443).
Safari and Firefox, however, demonstrates a fallback mechanism to port 443 if it fails to connect via the port suggested in the \https record.

\vspace{-1mm}
\paragraph{(2) IP hints.}
Browsers can bypass the additional \arec or \aaaa records lookups by leveraging the IP hints (\ipfh or \ipsh).
The \https record specification states that if local \arec/\aaaa records for \tname exist, clients should ignore these hints. 
Otherwise, clients are encouraged to conduct \arec and/or \aaaa queries for \tname and use the retrieved IP addresses (in the \arec/\aaaa records) for future connections. 
We interpret this as recommending clients prioritize IPs from \arec and \aaaa records and configure our DNS zone as below to test how browsers choose between these IPs.

\vspace{2mm}
\noindent
\fbox{\begin{minipage}{22em}
\begin{flushleft}
\small\texttt{a.com. 60 IN HTTPS 1 . alpn=h2 ipv4hint=1.2.3.4}\\
\small\texttt{a.com. 60 IN A 2.2.3.4} \\
\end{flushleft}
\end{minipage}}
\vspace{1.5mm}

\noindent As a result, we observe that both Safari and Firefox directly leverages the IP hints, while the Chrome and Edge prefer the IP addresses in \arec records. 

\paragraph{IP hints failover.}
To explore how browsers respond when they fail to establish connections with their preferred IP addresses, we set up servers exclusively accessible via either the IP address specified in the \ipfh or the one in the \arec record.
Safari makes an initial connection attempt with the first available IP address. If this attempt fails, it immediately retries with the IP address specified in the alternate record type. In contrast, Firefox waits for a longer period before attempting to connect with a different IP address. Chrome and Edge experience a hard failure if unable to connect using the IP addresses in the \arec record.

\paragraph{(3) \alpn.} 
To ensure successful access to the service endpoint, browsers should employ an application protocol advertised in the \alpn parameter.
We configure two server setups, each exclusively advertising either the HTTP/2 (\texttt{h2}) or HTTP/3 (\texttt{h3}) protocol.

\vspace{1.5mm}
\noindent
\fbox{\begin{minipage}{16em}
\begin{flushleft}
\small[1] \texttt{a.com. 60 IN HTTPS 1 . alpn=h2}\\
\small[2] \texttt{a.com. 60 IN HTTPS 1 . alpn=h3}\\
\end{flushleft}
\end{minipage}}
\vspace{1.5mm}

\noindent We observe that all four browsers successfully establish connections using the protocol specified in each \https record.
When the server exclusively advertises the HTTP/3 protocol, Firefox sends an HTTP/2 connection request shortly after initiating the correct protocol, possibly to ensure better compatibility.
When the server supports only HTTP/2, Firefox does not initiate an additional HTTP/3 connection.

\paragraph{Code corroboration.}
We examine the Chromium code and note that as of February 8th, 2024, it does not accommodate subsequent queries in \amode and \smode, indicating a lack of processing for follow-up queries for domain names in \tname beyond the apex. Moreover, Chrome does not initiate service on a separate port and disregards \https RR with an empty \alpn parameter. Furthermore, our investigation confirms that Firefox interprets \https parameters to establish connections in \smode. These findings are consistent with our experimental results.


\begin{tcolorbox}[width=\linewidth, sharp corners=all, colback=black!7, left=0.5pt, right=0.5pt, top=0.1pt, bottom=0.2pt]

\paragraph{(\textit{Takeaway})} 
Although major browsers query \https RR and fully support the \alpn parameter, they often fail to properly utilize other associated \https parameters. We identify several scenarios where inconsistent handling of \https record parameters can result in divergent connection behaviors.
In particular, such inconsistencies may direct connections to different IP addresses when there are mismatches between the \arec or \aaaa record and the IP hint parameter in the \https record, potentially leading to connection failures.

\end{tcolorbox}


\subsection{Browsers Support of ECH}
\label{subsec:browser_ech}
We now investigate browser support for ECH.
As discussed in \Cref{subsubsec:echdns}, managing ECH presents challenges due to its reliance on DNS.
Due to DNS caching, inconsistencies may arise between the ECH key (in the \https records) and the actual key used by servers.
Therefore, we examine how browsers respond to this issue, as well as other ECH misconfigurations.
Ultimately, our goal is to offer insights into challenges in ECH deployment, and help inform and improve future ECH deployment.\footnote{Recall that Cloudflare rolled back ECH deployment and has not re-enabled it at the time of this writing (\Cref{subsec:ech}).}




\paragraph{ECH workflow.}
We briefly describe the 
ECH workflow: 

\begin{packed_enumerate}
    \item To enable ECH, domain \texttt{private-example-ech.com.} publishes its ECH configuration via \https records, 
    including a public key for encrypting the \chello, an SNI extension directing to the client-facing server (that hosts \texttt{public-example-ech.com.}), 
    and other metadata.
    \item To access the target domain (i.e., \texttt{private-example-} \texttt{ech.com.}), a client fetches \https records, parses the ECH configuration (in the \ech parameter), and sends a \chello to the client-facing server (\texttt{public-example-ech}\texttt{.com.}). 
    This \chello includes a private \chello (termed \textit{ClientHelloInner}) that has an SNI pointing to the intended domain (i.e., \texttt{private-example-ech} \texttt{.com.}) 
    and is encrypted using the key advertised in \ech. 
    \item Upon receiving the \chello, the domain's client-facing server decrypts the encrypted \textit{ClientHelloInner} using its private key associated with the ECH configuration. 
    Successful decryption allows the client-facing server to forward the connection to the intended domain (i.e., \texttt{private-example-ech.com.} specified in an SNI of the \textit{ClientHelloInner}). 
    If decryption fails, the client-facing server either rejects and terminates the connection or sends ``retry ECH configurations'' to the client, which prompts the client to attempt the connection again using the newly provided ECH configuration.

\end{packed_enumerate}

\begin{figure}
    \centering
    \setlength{\abovecaptionskip}{2pt}
    \setlength{\belowcaptionskip}{-5pt}
    \includegraphics[width=.95\columnwidth]{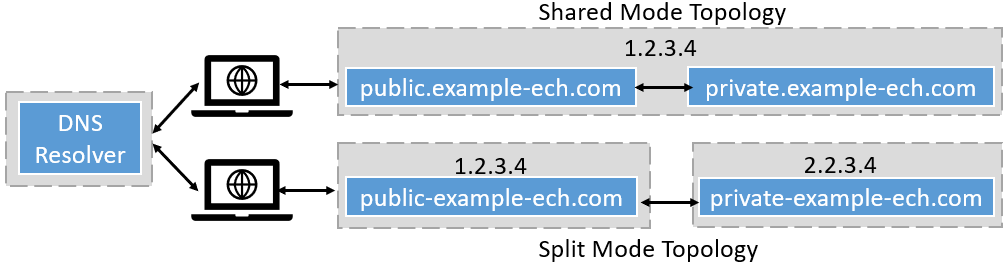}
    \caption{\texttt{Shared and} \spmode of ECH operation. 
    }
    \label{fig:ech-topology}
\end{figure}

There are two modes of ECH operation, \shmode and \spmode, as shown in \Cref{fig:ech-topology}). 
Each mode requires distinct configurations of \https records.
In \shmode, the client-facing server and the back-end server (e.g., web server) can be hosted on the same IP address, typically within the same apex zone (e.g., the same second-level domains). 
In \spmode, the client-facing and backend servers are hosted by a different apex zone and operate on separate IP addresses.
We will examine the support for both modes.

\subsubsection{\shmode ECH support} 
We set up \https records as follows.
The domain associated with the client-facing server is \texttt{cover.a.com.} and the domain associated with the back-end server is \texttt{a.com.}. Both domains' \arec records point to the same IP address (2.2.2.2).
The ECH configuration is specified in the \ech parameter.

\vspace{1.5mm}
\noindent
\fbox{\begin{minipage}{18em}
\begin{flushleft}
\small\texttt{a.com. 60 HTTPS 1 .  alpn=h2 ech=......}\\
\small\texttt{a.com. 60 A 2.2.2.2} \\
\small\texttt{cover.a.com. 60 A 2.2.2.2}
\end{flushleft}
\end{minipage}}
\vspace{1.5mm}

\noindent Upon directing the browsers to visit \texttt{https://a.com}, we note that three of the four browsers (except Safari) exhibit support for ECH,
by initiating a handshake with the client-facing server and encrypting the SNI in the \textit{ClientHelloInner}.

\begin{table}[!t]
\centering
\small
\begin{tabular}{r|c|c|c}
 & {\bf Chrome} & {\bf Edge} & {\bf Firefox} \\ \hline
\shmode Support & \fullcirc & \fullcirc & \fullcirc \\ \hline
(1) Unilateral ECH& \fullcirc & \fullcirc & \fullcirc \\ \hline
(2) Malformed ECH& \emptycirc & \emptycirc & \fullcirc \\ \hline
(3) Mismatched key  & \fullcirc &\fullcirc & \fullcirc  \\ \hline
\spmode Support & \emptycirc & \emptycirc & \emptycirc  \\ 
\end{tabular}
\caption{Browser support and failover mechanisms of ECH. 
 Same behavior is observed in Mac and Windows OS. Safari is excluded due to lack of any ECH support.}
\label{tab:ech-client}
\vspace{-8mm}
\end{table}

We next explore how browsers handle failover in the presence of misconfigured ECH through three experiments. 
Note that Safari, lacking ECH support, is omitted from our analysis. 
Our findings are summarized in \Cref{tab:ech-client}.

\paragraph{(1) Unilateral ECH deployment.}
In a scenario where a server no longer supports ECH but the associated domain's \https record continues to advertise ECH configuration, we examine the browsers' response.
This situation could arise if a server discontinues ECH support without updating its \https records to reflect this change.
Alternatively, even if the ECH configuration is removed from the domain's DNS zone file, clients might still attempt to connect using the cached ECH configuration.
Our findings indicate that three browsers successfully fallback to standard TLS connections.


\paragraph{(2) Malformed ECH configuration.}
We generate a malformed ECH configuration (e.g., due to typographical errors during copy-and-paste to zone files) that the browser cannot successfully parse. 
Chrome and Edge exhibit hard failure in the presence of malformed ECH configuration, terminating the connection after the initial SYN packet. 
This disrupts user access to the domain. 
In contrast, Firefox ignores the malformed ECH configuration and proceeds with a standard TLS handshake with the target server (i.e., \texttt{a.com}).

\paragraph{(3) ECH key mismatch.}
We publish a correct ECH configuration where the public key in the ECH diverges from the one utilized by the target server (i.e., the server that hosts \texttt{a.com}).
Such inconsistencies can arise from a failure to account for DNS cache effect in \https record management, as discussed in \Cref{subsubsec:echdns}.
The current ECH specification~\cite{ietf-tls-esni-17} outlines a server retry process (i.e., retry configuration) that can mitigate this problem.
This process entails the client-facing server offering a valid ECH configuration for retry, thereby allowing a client to reinitiate the TLS handshake with the valid configuration.
Our findings reveal that all three browsers support the retry mechanism and successfully establish connections with the target server (i.e., \texttt{a.com}) using the provided retry configuration. On the server-side, disabling retry is discouraged in the ECH specification~\cite{ietf-tls-esni-17} and is also not supported in the current ECH implementation of Nginx. We plan to further explore this aspect in future work.



\subsubsection{\spmode ECH Support}
In the \spmode topology, the client-facing and back-end servers may be hosted by separate entities, such as the ECH service provider and the website owner, across different apex zones and IP addresses. This introduces complexity in handling ECH configurations.

\vspace{1.5mm}
\noindent
\fbox{\begin{minipage}{21em}
\begin{flushleft}
\small\texttt{a.com. 60 HTTPS 1 . ech=..public\_name=b.com..}\\
\small\texttt{a.com. 60 A 1.1.1.1} \\
\small\texttt{b.com. 60 A 2.2.2.2} \\
\end{flushleft}
\end{minipage}}
\vspace{1.5mm}

\noindent Consider a scenario where the client-facing server, \texttt{b.com.} (specified as \texttt{public\_name} in the \ech parameter), operates on IP \texttt{2.2.2.2}, while the web server, \texttt{a.com.} (the domain the client intends to visit), is hosted on IP \texttt{1.1.1.1}.
To establish connections successfully, the client must interpret the ECH configuration, conduct subsequent DNS queries to locate the IP address of the client-facing server associated with \texttt{b.com.}, and then initiate a \chello to this server.

However, our experiments reveal that all three browsers fail to execute follow-up queries (i.e., \arec records for \texttt{b.com.}) and incorrectly initiate connections directly to the back-end server (i.e., 1.1.1.1) associated with \texttt{a.com.}, using the incorrect SNI of \texttt{b.com.}.
Consequently, the certificate validation process fails, leading to website loading failures across all three browsers; Chrome and Edge display an ``ERR\_ECH\_FALLBACK\_CERTIFICATE\_INVALID'' error, while Firefox shows a ``We're having trouble finding that site.'' message.

\begin{tcolorbox}[width=\linewidth, sharp corners=all, colback=black!7, left=0.5pt, right=0.5pt, top=0.1pt, bottom=0.2pt]

\paragraph{(\textit{Takeaway})} Although browsers (except for Safari) support ECH by default, certain essential features are still absent, especially in the \spmode where all three browsers hard fail on the connection. Such lack of support significantly harms servers with ECH configurations where their connectivity can be disrupted.

\end{tcolorbox}

%% file: tables/client-behavior.tex
\begin{table*}[!htbp]
\centering
\small
\begin{tabular}{c|c||cc|c|cc|cc}
 \multicolumn{2}{c||}{} & \multicolumn{2}{c|}{\textbf{Chrome}} & \textbf{Safari} & \multicolumn{2}{c|}{\textbf{Edge}} & \multicolumn{2}{c}{\textbf{Firefox}} \\ 
  \specialrule{.11em}{.1em}{.1em}
 
 \multicolumn{2}{c||}{\textbf{OS}} & \multicolumn{1}{c|}{macOS} & Windows & macOS & \multicolumn{1}{c|}{macOS} & Windows & \multicolumn{1}{c|}{macOS} & Windows \\ \hline
 
 \multicolumn{2}{c||}{\textbf{Browser Version}} & \multicolumn{2}{c|}{120.0.6099}  & 17.2.1 & \multicolumn{2}{c|}{120.0.2210} & \multicolumn{2}{c}{122.0.1} \\ 
 \specialrule{.11em}{.1em}{.1em}
\multirow{3}{*}{\shortstack[c]{\bf{\https RR} \\ \bf Utilization}} & \textbf{\{apex\}} & \multicolumn{1}{c|}{\fullcirc} & \fullcirc & \halfcirc & \multicolumn{1}{c|}{\fullcirc} & \fullcirc & \multicolumn{1}{c|}{\fullcirc} & \fullcirc \\ \cline{2-9} 

 & \textbf{http://\{apex\}} & \multicolumn{1}{c|}{\fullcirc} & \fullcirc & \halfcirc & \multicolumn{1}{c|}{\fullcirc} & \fullcirc & \multicolumn{1}{c|}{\fullcirc} & \fullcirc \\ \cline{2-9} 
 
 & \textbf{https://\{apex\}} & \multicolumn{1}{c|}{\fullcirc} & \fullcirc & \fullcirc & \multicolumn{1}{c|}{\fullcirc} & \fullcirc & \multicolumn{1}{c|}{\fullcirc} & \fullcirc \\ \hline\hline
 
 
 \multirow{1}{*}{\textbf{\amode}} & \textbf{\tname} & \multicolumn{1}{c|}{\emptycirc} & \emptycirc & \fullcirc & \multicolumn{1}{c|}{\emptycirc} & \emptycirc & \multicolumn{1}{c|}{\emptycirc} & \emptycirc \\ \hline \hline
 
 \multirow{4}{*}{\textbf{\smode}} & \textbf{\tname} & \multicolumn{1}{c|}{\emptycirc} & \emptycirc & \fullcirc & \multicolumn{1}{c|}{\emptycirc} & \emptycirc & \multicolumn{1}{c|}{\fullcirc} & \fullcirc \\ \cline{2-9} 
 & \textbf{\port} & \multicolumn{1}{c|}{\emptycirc} & \emptycirc & \fullcirc & \multicolumn{1}{c|}{\emptycirc} & \emptycirc & \multicolumn{1}{c|}{\fullcirc} & \fullcirc \\ \cline{2-9} 
 & \textbf{\alpn} & \multicolumn{1}{c|}{\fullcirc} & \fullcirc & \fullcirc & \multicolumn{1}{c|}{\fullcirc} & \fullcirc & \multicolumn{1}{c|}{\fullcirc} & \fullcirc \\ \cline{2-9}  
 & \textbf{IP Hints} & \multicolumn{1}{c|}{\emptycirc} & \emptycirc & \fullcirc & \multicolumn{1}{c|}{\emptycirc} & \emptycirc & \multicolumn{1}{c|}{\fullcirc} & \fullcirc \\  

\end{tabular}
\caption{The \https RR support from four major browsers. 
A full circle (\fullcirc) means that the record or parameter being utilized. A half circle (\halfcirc) suggests that while the record or the parameter is being utilized, some essential function related to it is missing. An empty circle (\emptycirc) denotes that there is no support for the feature.}
\label{tab:client-https}
\vspace{-6mm}
\end{table*}

%% file: sections/7_relatedwork.tex
\section{Related Work}
\label{sec:relatedwork}


\paragraph{\https RR Deployment.} 
While there are numerous studies on the deployment of DNS record types (e.g., DNSSEC records), \https (as well as \svcb) records have received scant attention, due to their recent introduction.
There has been research~\cite{zirngibl2021s} primarily examining the interaction between \https records and \texttt{QUIC} deployment. 
However, this study used \https records as a means to analyze the deployment of the \texttt{QUIC} protocol, rather than performing an analysis of the \https records ecosystem.
%
%
In 2023, Zirngibl and colleagues~\cite{zirngibl2023first} performed scans on over 400 million domains within a 15-day timeframe to examine the deployment of \svcb and \https records.
They uncovered that about ten million domains support \https records, with a majority hosted by \textit{Cloudflare}.
Jan Schaumann~\cite{jschauma-https-deploy} released a brief analysis on the adoption and usage of \https RRs in Tranco top 1M domains.
%

Our study diverges from these prior works in several ways.
First, to the best of our knowledge, this is the first study to dissect browser support for \https RR, including their failover mechanisms.
Furthermore, we examine how browsers handle \ech in various configurations (including misconfigurations).
Second, although our datasets cover a smaller number of domains compared to ~\cite{zirngibl2023first}, our focus extends to the longitudinal analysis of \https records, including inconsistent use of \https records, changes in DNS providers, IP address inconsistency, and domain connectivity. 
We also perform in-depth analysis to understand the implications of \https parameters through additional experiments, such as the connectivity experiments to examine the reachability of mismatched IPs, additional scans to measure the \ech key rotation frequency, and the conjunction with DNSSEC to understand the security protection. Last but not least, our analysis distinguishes between popular domains (overlapping) and relatively less popular domains (dynamic Tranco). This approach allows us to explore trends based on domain popularity more thoroughly.

\paragraph{ECH Adoption.} 
Several studies investigated ECH.
Chai \textit{et al.}~\cite{chai2019importance} examined Encrypted SNI (ESNI, a precursor to ECH) in the context of censorship circumvention. 
Bhargavan \textit{et al.}~\cite{bhargavan2022symbolic} conducted an assessment of the security, privacy, and performance aspects of TLS 1.3 with and without ECH by employing automated verification tools. 
Tsiatsikas \textit{et al.}~\cite{tsiatsikas2022measuring} analyzed the adoption rates of both ESNI and ECH 
but found only one domain that supported ECH\@.
Most recently, Zirngibl \textit{et al.}~\cite{zirngibl2023first} reported 20 domains utilizing ECH among 400 millions domains scanned in 2023. 

We observed a marked increase in the usage of ECH as compared to previous research. 
Our longitudinal study indicates that \cf began adopting ECH as early as May 2023, four months prior to their announcement~\cite{cloudflare-ech}. 
Furthermore, while earlier studies do not address the service providers associated with ECH, our study not only examines these providers but also explores ECH key rotation and its integration with DNSSEC.
Additionally, we assess ECH support across popular web browsers. 



%% file: sections/8_1_discussion.tex
\section{Discussion}
\label{sec:discussion}

\paragraph{Automation tool for \https record management.}
Managing \https records involves several complexities due to their DNS-based nature, including the coordination across multiple DNS service providers, potential inconsistencies between IP hints and \arec/\aaaa records, and the risks associated with improper handling of ECH, which can lead to connection issues.
We believe the DNS \https ecosystem could borrow experiences learned from the management of digital certificates, 
where automating the certificate issuance and renewal process through \texttt{ACME} and \texttt{Certbot}~\cite{aas2019let,tiefenau2019usability} has significantly reduced the barriers to obtaining and maintaining a digital certificate, demonstrating the potential benefits of automation in managing web security features.

\paragraph{Limitations in major browsers.}
Our assessment indicates that all four leading web browsers currently support querying \https records, which points to a promising trend in industry adoption. However, several crucial functionalities remain unimplemented. Notably, both Chrome and Edge lack support for IP hints—a parameter utilized by 97\% of apex domains and 87\% of www domains that have adopted \https records. Furthermore, the absence of support for ECH \texttt{Split Mode} could potentially lead to service disruptions. Cloud providers and domain administrators should take these limitations into account when integrating \https records into their systems.

%% file: sections/8_2_conclusion.tex
\section{Conclusion}
\label{sec:conclusion}

In this study, we present a comprehensive analysis of the DNS \https record ecosystem, uncovering the deployment challenges and complexities from both server-side and client-side perspectives. 
Specifically, our server-side analysis shows that over 20\% of domains in the Tranco list support \https records, with \cf playing a crucial role in this adoption and a noticeable increase in support from other major DNS providers as well. However, a significant concern is the lack of DNSSEC protection for many \https records, particularly those utilizing ECH, which renders them vulnerable to potential attacks. 
We also explore the complexities of managing \https records, including issues related to IP hints and ECH configurations. 
On the client side, while the four major web browsers support \https record lookups, they do not fully utilize the capabilities offered by \https records. 
Our analysis reveals that improper handling of \https records can lead to connection failures, shedding light on the obstacles that we need to overcome to move towards a more widespread \https deployment. We plan to reach out to DNS providers and web browsers regarding our findings.

%% file: sections/appendix.tex
\section{Ethics}
\label{appendix:ethics}
Throughout our scanning activities, we strictly adhered to ethical standards~\cite{dittrich2012menlo,partridge2016ethical}. 
Our study potentially impacts two entities: DNS resolvers and authoritative name servers hosting Tranco Top 1M domains. To mitigate any negative effects, we take the following precautions.
First, we conduct our scans at a controlled pace to ensure that we never overwhelm a single resolver with numerous concurrent requests simultaneously.  
Next, we limit our data retrieval to only the necessary DNS records for our analysis (as outlined in \Cref{tab:server_dataset}), which we collect once daily.
For specific analyses, we conduct additional scans (e.g., for DNSSEC records).
We consider the load placed on the name servers of domains within the Tranco list due to our DNS scans is negligible, especially considering their popularity (and high levels of traffic).
Additionally, we clearly identified our measurement vantage point through DNS and WHOIS information, and a maintained testbed including a hosted domain and an operated authoritative name server using BIND9. 
Notable, we did not encounter any inquiries regarding our scans throughout this endeavor.

\section{Background}
\label{appendix:background}

\paragraph{DNS record types.}
DNS records, known as resource records, are entries allowed in DNS zone files that serve to associate domain names with IP addresses and offer additional information about domains.
We introduce DNS record types used in this work as follows (\https records, the main focus of this study, will be detailed in the subsequent paragraph):
\begin{packed_itemize}
    \item \arec maps a domain name to its IPv4 address. 
    \item \aaaa maps a domain name to its IPv6 address.
    \item \cname (\texttt{Canonical Name}) creates an alias from one domain name to another. When a DNS resolver encounters a \cname record, it replaces the original domain name with the canonical domain name specified in the record and then performs a new DNS lookup using the canonical name.
    \item \soa (\texttt{Start of Authority}) holds information about a DNS zone. This record is crucial for managing DNS zones and ensuring the proper functioning of DNS services. 
    \item \ns (\texttt{Name Server}) stores information of the authoritative name servers for a domain.
    \item \rrsig (\texttt{Resource Record Signature}) stores cryptographic signatures of DNS records, used to authenticate records in accordance with DNSSEC (DNS Security Extensions)~\cite{rfc4033,rfc4034,rfc4035}. If the \rrsig record passes validation, the integrity of the given DNS record is ensured.
    \item \dnskey (\texttt{DNS Public Key}) holds a cryptographic public key used to verify an \rrsig record (of a given DNS record). 
    \item \ds (\texttt{Delegation Signer}) contains the hash of a key (in \dnskey) and is uploaded to the parent DNS zone to create a chain of trust across the DNS hierarchy.
\end{packed_itemize}

\begin{figure}[h]
 \centering
    \setlength{\abovecaptionskip}{2pt}
    \includegraphics[scale=.4]{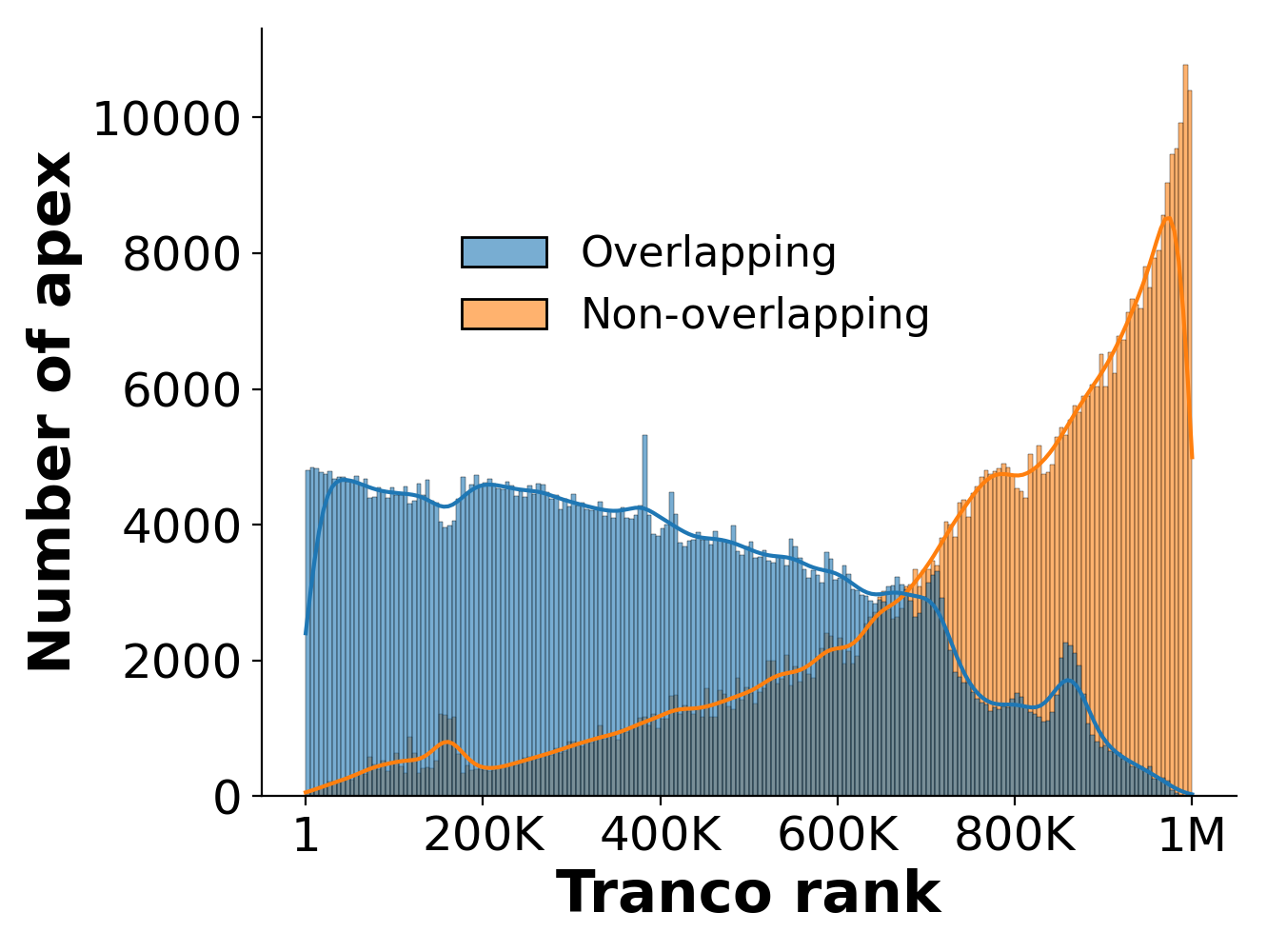}
    \caption{The distributions of Tranco rankings for each group of apex domains (overlapping or non-overlapping). The rank of each domain is averaged over the period from May 8th, 2023 to July 31st, 2023.}
    \label{fig:apexranking}
\vspace{-4mm}
\end{figure}

\section{Domains in the Tranco List}
\label{appendix:tranco}
Given the daily updates to the Tranco list, the list's domain composition can change daily.
Therefore, relatively popular domains (e.g., those with higher rankings) are likely to consistently appear in the list, while less popular domains (i.e., those with lower rankings) may not consistently included in the list.
\Cref{fig:apexranking} shows the distribution of average popularity rankings for (apex) domains that are consistently included in the Tranco list throughout the entire first measurement period (i.e., prior to the list's source change), compared to those that are not; say overlapping and non-overlapping domains, respectively.

\begin{figure}[h]
 \centering
    \setlength{\abovecaptionskip}{2pt}
    \includegraphics[scale=.35]{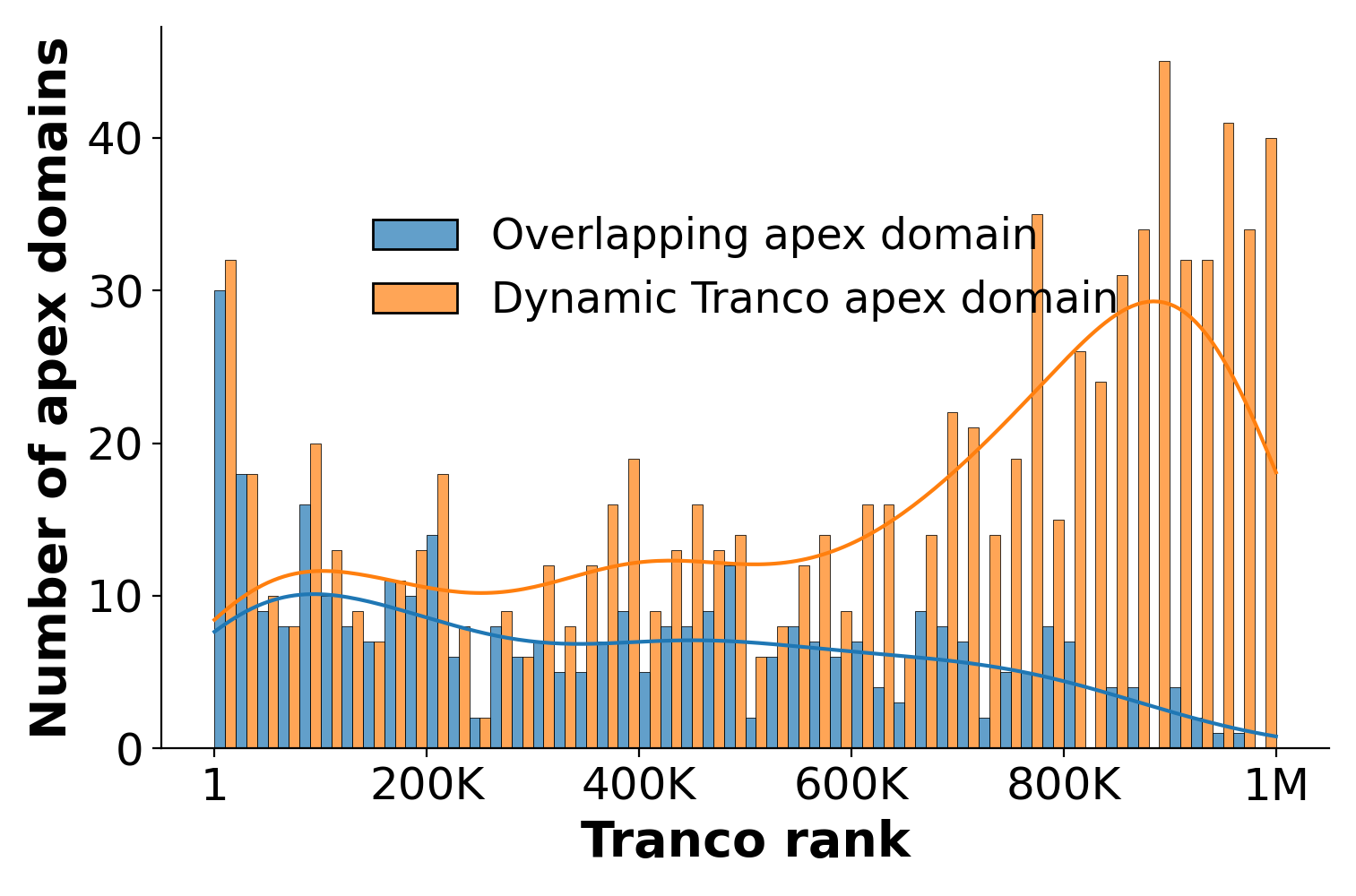}
    \caption{Ranking of apex using non-\textit{Cloudflare} name servers. The rank of each apex is shown by its mean ranking value across the period from Oct 11th, 2023, to Jan 21st, 2024. }
    \label{fig:otherns-apexrank}
\vspace{-4mm}
\end{figure}

We observe that the overlapping domains tend to include domains with higher rankings compared to non-overlapping domains.

\section{Domains with Non-\cf Name Servers}
\label{appendix:noncf}

\subsection{Ranking of Domains with Non-\textit{Cloudflare} Name Servers}
\label{appendix:ranking-noncf}

We show the ranking of these apex domains in Figure~\ref{fig:otherns-apexrank}.

\subsection{Domains with HTTPS records}
\label{appendix:domain-noncf-httpsrr}

\begin{figure}[!ht]
 \centering
    \setlength{\abovecaptionskip}{2pt}
    \includegraphics[scale=.20]{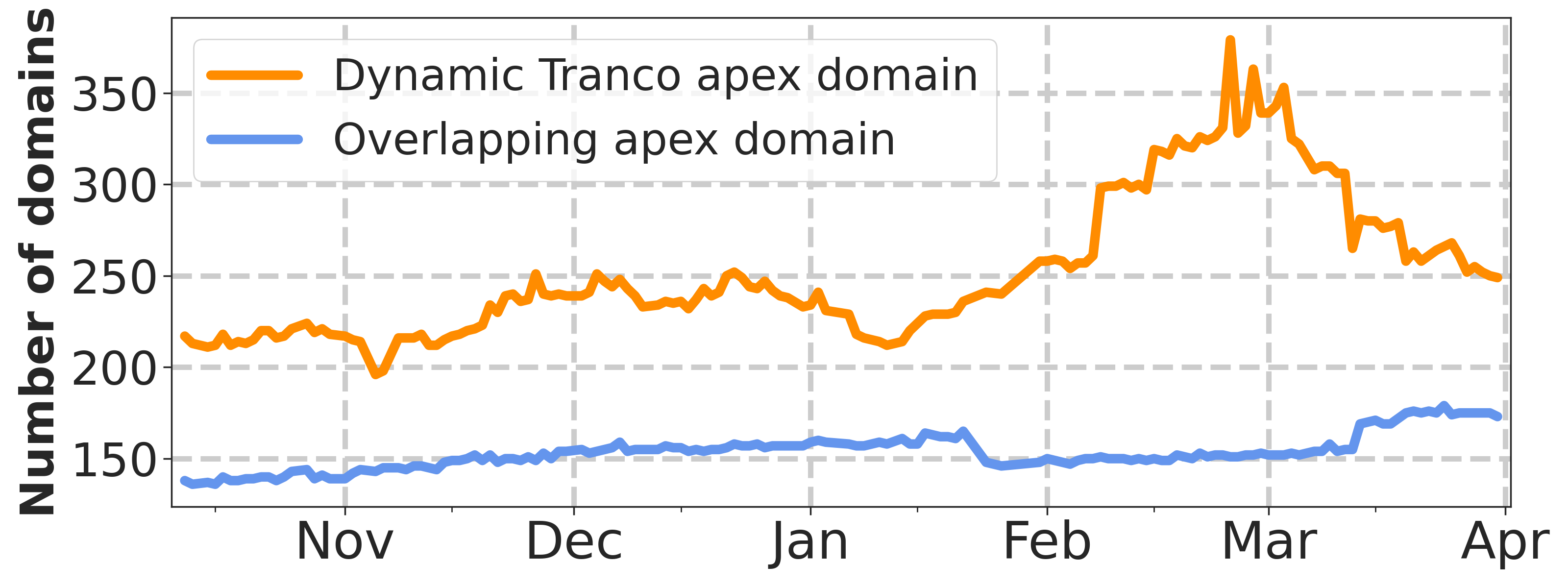}
    \caption{Number of domains that both activate \https records and use non-\cf name servers.}
    \label{fig:otherns-apex}
\end{figure}

We show the number of apex domains utilizing non-\cf name servers during \https record activation in Figure~\ref{fig:otherns-apex}.

\section{Details on \https RR Parameters}
\label{appendix:https-parameters}

We present additional details of the \https RR parameters. 


\subsection{\sprty and \tname}
\label{appendix:param_target}

During our measurement period, the \sprty value of 1 (i.e., \smode) is adopted by 99.97\% and 99.95\% of \https RR on average, for overlapping apex domains and \www subdomains respectively. 
On the other hand, the \sprty value of 0 (\amode), meanwhile, is used by approximately 39 \https records for apex domains and 7 for \www subdomains, on a daily average, respectively. 

The dominance of \smode, indicated by the value 1, is largely due to domains utilizing \textit{Cloudflare} name servers with the default \https record configurations, as we discussed in the previous section (\Cref{subsubsec:cloudflare}). 

\paragraph{Domains with \textit{Cloudflare} name servers.}
We observe that on a daily average, approximately 12 apex domains using \textit{Cloudflare} name servers have customized \https record configuration (i.e., setting their \sprty or \tname differently from \textit{Cloudflare}'s default configuration). 
Among these domains, 5 apex domains use \sprty value of 0 (i.e., \amode). 
However, one domain, \texttt{newlinesmag.com}, sets itself as the \tname (i.e., by using ``.'' as value), despite using the \amode.
Other three apex domains (\texttt{unze.com.pk}, \texttt{idaillinois.org}, and \texttt{pokemon-arena.net}) diverge from standard practices by using IP addresses as their \tname. 
Lastly, \texttt{gachoiphungluan.com} uses an \https URL as its \tname.
Additionally, we note that 14 distinct apex domains specify the same \tname,  \texttt{geo-routing.nexuspipe.com}, with multiple \sprty values in their corresponding \https RR. 
The \sprty values for all these \https records are a list including values ranging from 1 to 12, with each corresponding to a specific port. 
Interestingly, domain \texttt{host-ir.com} and \texttt{pionerfm.ru} keep only one \https RR and they use priority 443 and 1800, respectively, for the record.

\begin{table}[t]
\vspace{-2mm}
\centering
\small
\begin{tabular}{ll|cc}
\multicolumn{2}{c|}{\multirow{3}{*}{\bf Protocols}}                                              & \multicolumn{2}{c}{\multirow{2}{*}{\shortstack[c]{\bf{\% of domains} \\ \bf{with \https RR}}}} \\
\multicolumn{2}{l|}{}                                                                                & \multicolumn{2}{c}{}                                                      \\ \cline{3-4} 
\multicolumn{2}{l|}{}                                                                                & \multicolumn{1}{c|}{\textbf{Apex}}     & \textbf{\www}     \\ \hline
\multicolumn{2}{l|}{\textbf{\textcolor{red}{HTTP/2}}}                                                                 & \multicolumn{1}{c|}{99.64}                    & 99.61                     \\ \hline
\multicolumn{2}{l|}{\textbf{\textcolor{red}{HTTP/3}}}                                                                 & \multicolumn{1}{c|}{78.42}                    & 75.67                      \\ \hline
\multicolumn{1}{l|}{\multirow{2}{*}{\textbf{HTTP/3-29}}} & $<$ May 31st, 2023       & \multicolumn{1}{c|}{77.43}                    & 74.32                      \\ \cline{2-4} 
\multicolumn{1}{l|}{}                                      & $\geq$ May 31st, 2023 & \multicolumn{1}{c|}{$<$ 0.01}           & $<$ 0.01             \\ \hline
\multicolumn{2}{l|}{\textbf{HTTP/3-27}}                                                            & \multicolumn{1}{c|}{$<$ 0.01}           & 0                          \\ \hline
\multicolumn{2}{l|}{\textbf{HTTP/1.1}}                                                               & \multicolumn{1}{c|}{$<$ 0.01}           & $<$ 0.01             \\ 
\end{tabular}
\caption{Application layer protocols specified in the \texttt{alpn} parameter of overlapping domains, on a daily average. Protocols highlighted in red are those in \cf's default \https record configuration.}
\label{tab:alpn}
\vspace{-8mm}
\end{table}

\paragraph{Domains with non-\textit{Cloudflare} name servers.}
When examining apex domains that utilize non-\textit{Cloudflare} name servers, we discover 2,884 such domains, with 2,755 (95.53\%) of them using a \sprty value of 1 (i.e., \smode) and setting their \tname to point to themselves (by using the value of ``.''). 
We observe 9 domains using a \sprty value of 1 (i.e., \smode) but setting their \tname to point to alternatives. 
There are 108 domains that utilize the \amode (with \sprty value of 0), among which 22 apex domains set themselves as the \tname; note that 21 out of 22 apex domains are with \texttt{domaincontrol.com} name servers and 1 employ \{\texttt{he.net}, \texttt{shaunc.com}, and \texttt{shat.net}\} as its name server hosts. 
Additionally, we again observe 7 domains with a list of priority values using non-\textit{Cloudflare} name servers. These domains also specify the same \tname (\texttt{geo-routing.nexuspipe.com}) with \sprty ranging from 1 to 12, with each assigned to a specific port. These domains are using \texttt{sone.net} name servers.  

While \sparam is an optional field reserved for \smode, our observation reveals that the majority of domains employing \smode include at least one key-value pair.
In contrast, 232 apex domains utilize \sprty value of 1 (i.e., \smode) but do not provide any \sparam.
This encompasses 42 DNS service providers, with the most prevalent being \texttt{google.com}, \texttt{domaincontrol.com}, \texttt{netclient.no}, \texttt{icsn.com}, \texttt{nsone.net}, \{\texttt{d-53.jp}, \texttt{d-53.net}, and \texttt{d-53.info}\}. 
The results for \www domains are largely similar to the apex domains.


\subsection{ALPN}
\label{appendix:alpn}

The prevalent protocols in \https RR are HTTP/2 and HTTP/3, garnering support from almost 100\% for overlapping domains (and near 80\% for \www subdomains), as shown in \Cref{tab:alpn}. 
These substantial percentages align with our earlier observation that the majority of apex domains employ \cf name servers and maintain an \https RR configuration identical to \cf's default settings (see Section~\ref{subsubsec:ns}).

It is noteworthy that we observe massive support of the implemented draft version 29 of HTTP/3 prior to May 31st, 2023; starting from May 31st, 2023, we merely observe several support of this draft version on a daily basis. 
This aligns with the fact that as of late May 2023, \textit{Cloudflare} no longer advertises this draft version for zones that have HTTP/3 enabled~\cite{cloudflare-alpn}. 

Among overlapping apex domains employing \textit{Cloudflare} name servers with customized \https records, HTTP/2 is supported by approximately 98.57\% domains on a daily basis.
In contrast to domains with \textit{Cloudflare}'s default \https record configurations, only 0.28\% advertise their support for HTTP/3, and 1.13\% do not include \alpn in their \sparam.

For apex domains utilizing non-\cf name servers, we observe a lower ratio of advertising HTTP/2 and HTTP/3 as compared to \cf's default configuration, with an average of 64.09\% and 26.79\%, respectively. 
About 8.44\% domains do not include \alpn in their \https records. 
Moreover, we observe 1 domains continuously advertise both draft version 27 and 29 of HTTP/3. 
This apex domain \texttt{gentoo.org} is with the \texttt{gentoo.org} name server, suggesting that it hosts its own apex zone.  
Specifically, we observe 6 apex domains exclusively advertise HTTP/1.1, of which 2 utilize a combination of name server \texttt{jpberlin.de} and \texttt{cloudns.net}, 2 employ \texttt{jpberlin.de}, 1 use a mix of \texttt{gandi.net} and \texttt{trash.net}, and the remaining 1 hosts it own apex zone.\footnote{\texttt{he.net}, \texttt{shaunc.com}, \texttt{shat.net}.} 

Among \www subdomains, 1.63\% and 0.18\% do not indicate support for any \texttt{alpn}, for subdomains with customized \https records using \cf name servers and with non-\cf name servers, respectively.

Additionally, starting from Feb 11th, 2024, we observe a consistent 0.003\% of domains supporting \textit{Google} \texttt{QUIC} version Q043, Q046, and Q050. These domains are all with \cf name servers.


\subsection{IP Hint Mismatching Analysis}
\label{appendix:iphint}

\begin{figure}[!h]
    \centering
    \setlength{\abovecaptionskip}{5pt}
    \setlength{\belowcaptionskip}{1pt}
    \begin{subfigure}[b]{.99\columnwidth}
        \includegraphics[width=.97\columnwidth]{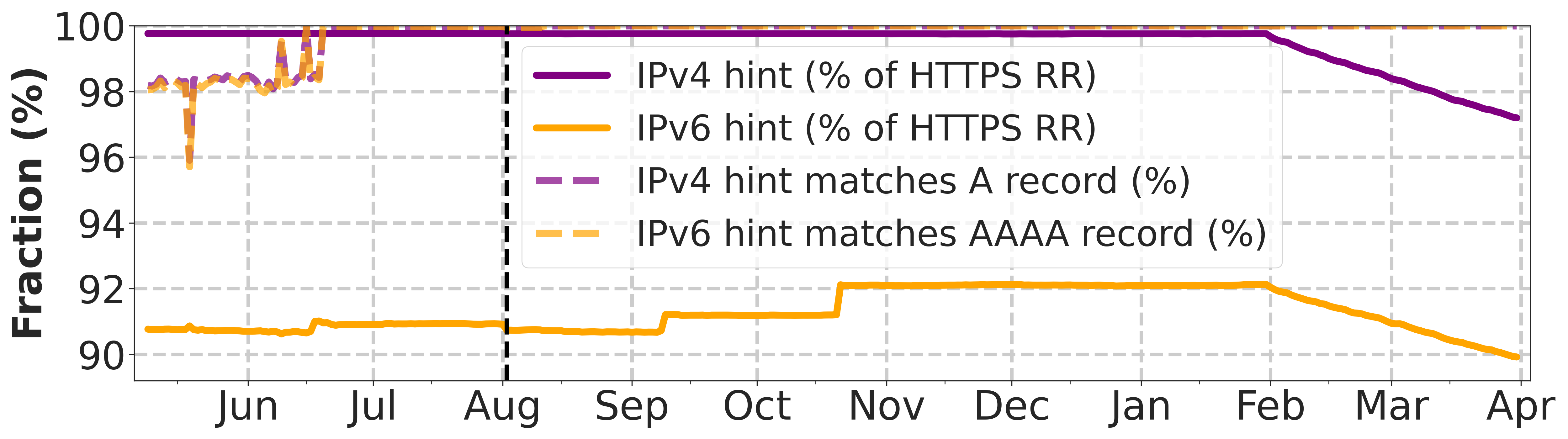}
        \caption{IP hints utilization and consistency with \arec/\aaaa records in overlapping apex domains.}
        \label{fig:iphint-apex}
    \end{subfigure}
    \begin{subfigure}[b]{.99\columnwidth}
        \includegraphics[width=.97\columnwidth]{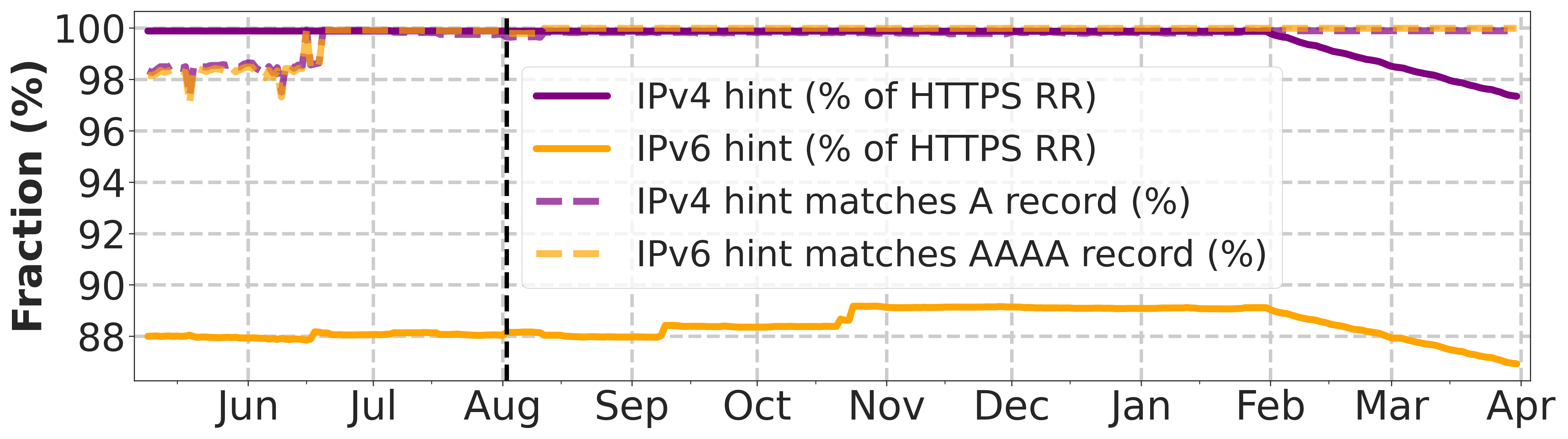}
        \caption{IP hints utilization and consistency with \arec/\aaaa records in overlapping \www subdomains.}
        \label{fig:iphintwww}
    \end{subfigure}
    \caption{The ratio of domains with \https records that utilize IP hints (solid lines) and their matching ratio with IP addresses in \arec/\aaaa records (dashed lines). The vertical dashed line (August 1st, 2023) denotes the source change of the Tranco list.}
    \label{fig:iphint}
\vspace{-4mm}
\end{figure}

Figure~\ref{fig:iphint} show the ratio of overlapping domains that specify \\
\ipfh/\ipsh in their \https records (solid lines), as well as the consistency between the IP addresses provided in the IP hints and those in the corresponding \arec/\aaaa records of the domains (dashed lines).
Before June 19th, 2023, the matching rates fluctuated around 98\% for both apex domains and \www subdomains.
However, starting from June 19th, 2023, the matching rates (for both apex domains and \www subdomains) increased to over 99.8\% for \ipfh and \ipsh, aligning with the corresponding \arec/\aaaa records.

To take a closer look into the mismatched IP addresses in IP hints v.s. the corresponding \arec/\aaaa records (\Cref{fig:iphint}), 
we examine name servers utilized by these domains.
Unfortunately, we lack information on name servers utilized by domains before August 16th, 2023 (as our \ns records scan started on this date).
To address this gap, we estimate their name servers by cross-referencing the name servers these domains utilized after August 16th, 2023.
Based on this approach, we then continue our analysis on name servers.


\paragraph{IP hints and name servers with cross-referencing.}
Before June 19th, 2023, 40,578 of apex domains and 36,825 \www domains (both about 2\% of domains utilizing \https RR) exhibit inconsistencies between their IP hints and \arec/\aaaa records. 
As a result, we can estimate the name server for 88.08\% of apex domains and 86.46\% of \www subdomains before June 19th, 2023.
We observe that 99.97\% of domains with mismatches between their IP hints and \arec/\aaaa records utilize \cf name servers; the remaining domains use \texttt{cf-ns.com} and \texttt{cf-ns.net} name servers.

Among these domains, about 64\% exhibit inconsistency in both \ipfh and \ipsh (for both apex domains and \www subdomains). 
While the majority of these domains display inconsistency in just a few days, 
we observe that 14 apex domains and 17 \www subdomains consistently show such discrepancies for over 10 days. 
Furthermore, 5 apex domains and 8 \www subdomains consistently present mismatched IP hints and \arec/\aaaa records throughout the entire observation period, all of which are associated with \cf name servers.

After the matching rate increase on June 19th, 2023, we identify discrepancies in IP hints and \arec/\aaaa records for 178 apex domains and 814 \www subdomains, through the cross-referencing of name servers, with daily discrepancies ranging from 30 to 80 domains.
Among these, 94.94\% and 98.89\% are associated with \textit{Cloudflare} name servers, while the remainder utilize \texttt{cf-ns.com}, \texttt{cf-ns.net}, \texttt{peavey.com}, and \texttt{upclick.com} name servers.
Interestingly, 52 apex domains and 116 \www subdomains are also found to advertise mismatched IP addresses before June 16th, 2023. 
Once again, the majority of these domains use \textit{Cloudflare} name servers, with the rest employing \texttt{cf-ns.com} and \texttt{cf-ns.net} name servers.



\paragraph{IP hints and name servers after August 16th, 2023.}
Since we start scanning the name servers of domains on August 16th, 2023, we directly utilize the collected data for our analysis.
We observe a total of 482 apex domains and 4,508 \www subdomains advertising mismatched IP addresses, involving 91 and 34 DNS service providers, respectively. 
Notably, while \texttt{cf-ns} name servers are predominantly utilized by apex domains exhibiting such inconsistencies, \www subdomains with mismatched IP addresses primarily employ \textit{Cloudflare} name servers.
Furthermore, we identify 4 apex domains and 4 \www subdomains consistently advertising mismatched IP addresses from May 8th, 2023, to January 21st, 2024; all are associated with \texttt{cf-ns} name servers.

\begin{figure}[t]
\vspace{-3mm}
    \centering
    \setlength{\abovecaptionskip}{0pt}
    \setlength{\belowcaptionskip}{-5pt}
    \includegraphics[width=.8\columnwidth]{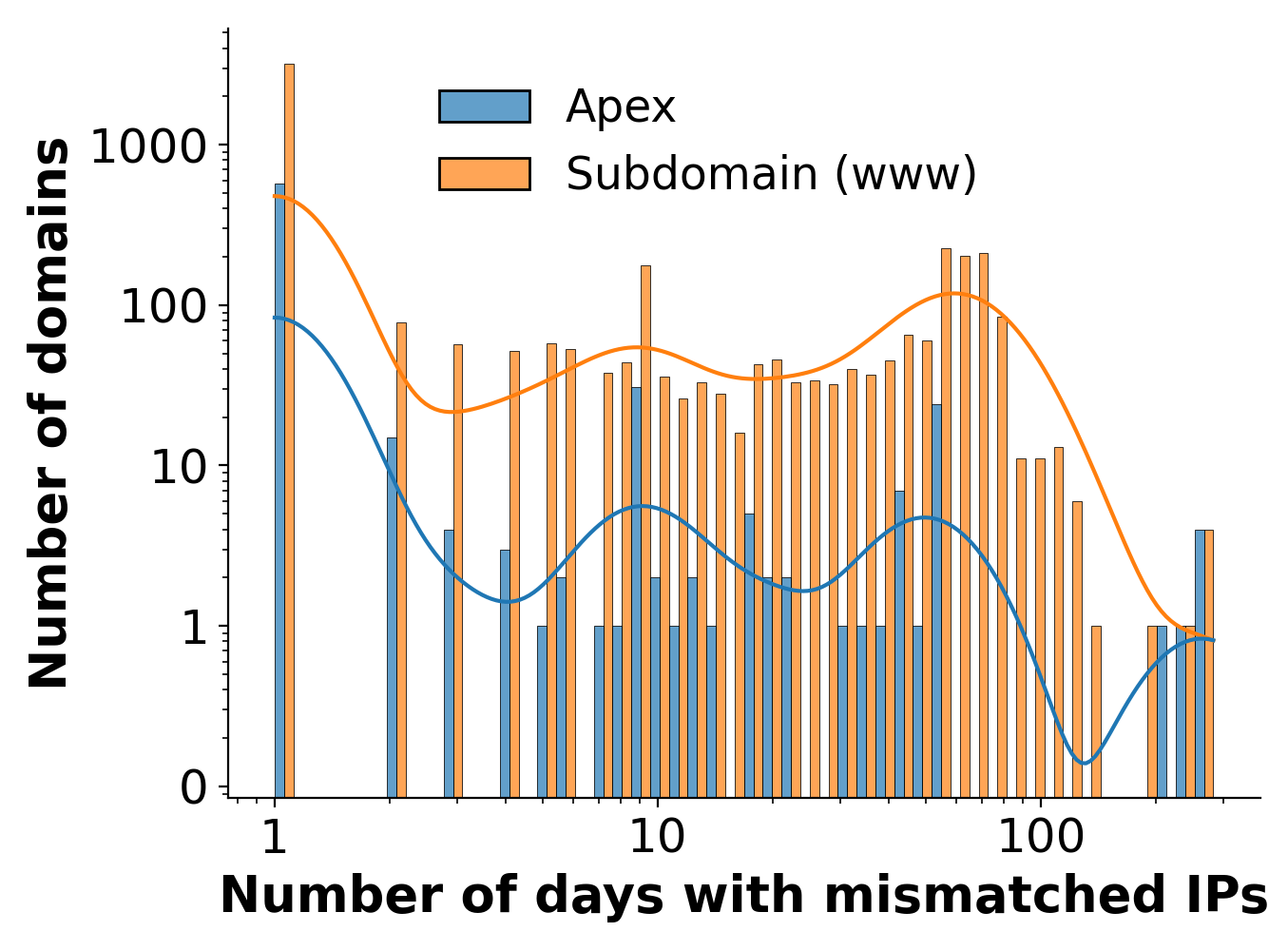}
    \caption{Duration of domains with mismatched IP hints and A/AAAA records.}
    \label{fig:ipmismatch-dur}
\vspace{-6mm}
\end{figure}

\paragraph{Mismatch duration.} 
We monitor domains exhibiting discrepancies between IP hints and A/AAAA records commencing on June 19th, 2023, and illustrate the duration of these mismatches in Figure~\ref{fig:ipmismatch-dur}. 
In particular, we find 4 apex domains and 4 \www subdomains that consistently provide mismatched IP addresses throughout our data collection period from May 2023 to March 2024; 
among these, 66.67\% and 93.22\% are associated with \texttt{cf-ns} (name servers for \cf China Network, partnered with Chinese registrars~\cite{cf-ns}) and \textit{Cloudflare} name servers, respectively.

\section{ECH Deployment by Domains}
\label{appendix:ech_deployment}

\begin{figure}[t]
    \centering
    \setlength{\abovecaptionskip}{5pt}
    \setlength{\belowcaptionskip}{1pt}
    \includegraphics[width=.97\columnwidth]{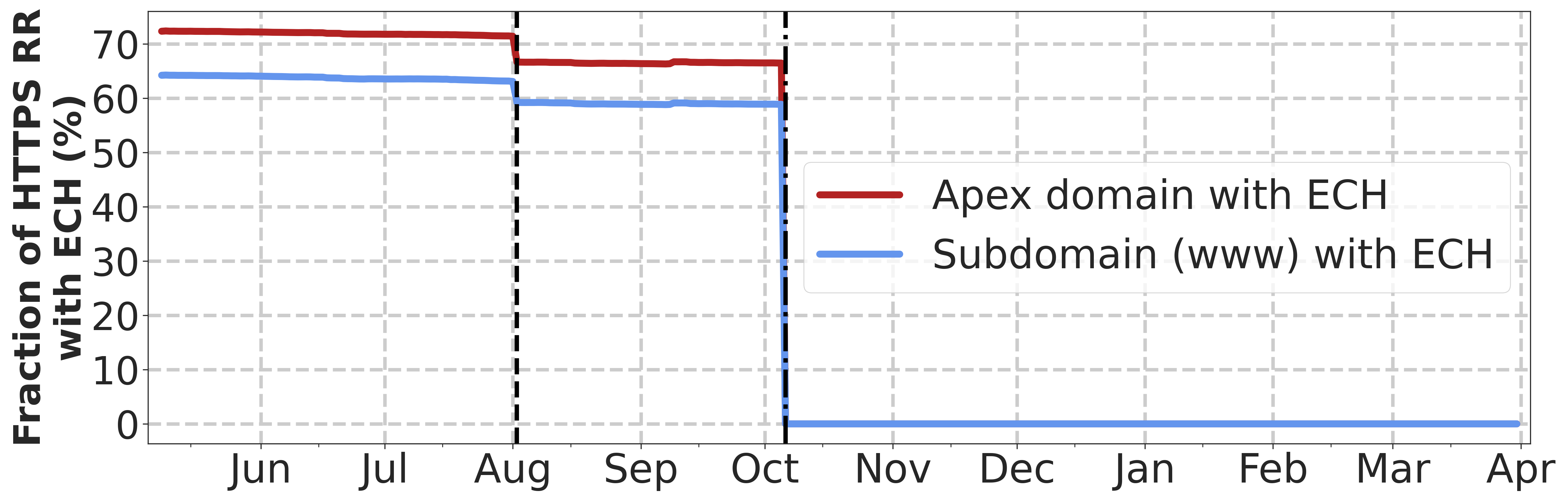}
    \caption{The percentage of overlapping domains with \https records that support ECH. The vertical dashed line (near August 1st, 2023) denotes the source change of the Tranco list, and the vertical dash-dotted line (on October 5th, 2023) shows the date that \cf disabled ECH from all its domains.}
    \label{fig:ech}
\vspace{-8pt}
\end{figure}

\Cref{fig:ech} illustrates the ratio of domains that have deployed ECH (by publishing \ech parameters) among those that publish \https records.


\section{DNSSEC Analysis of Apex Domains and Name Servers}
\label{appendix:dnssec-apex}

\paragraph{Comparison with domains without \https RR.}
We perform an additional data collection on January 2, 2024, 
where we fetch and validate the DNSSEC chain (i.e., \dnskey, \ds, and \rrsig records) of top 1M apex domains, using the Unbound library~\cite{unbound}.
\Cref{tab:dnssec-stats} shows the number of domains with signed records and their corresponding DNSSEC validation results.
Interestingly, we find that 49.4\% of signed \https records are \insecure (i.e., missing the \ds records in their parent zone)~\cite{rfc4033}.
This ratio is considerably high compared to the commonly known \insecure ratio of domains that support DNSSEC; in our data, only 23.7\% of signed domains (i.e., have a \dnskey record) that do not publish \https records are \insecure (as indicated in the first row of~\Cref{tab:dnssec-stats}), and this finding aligns with the similar \insecure ratio of around 30\% reported in \cite{chung2017longitudinal}.
We also examine the \insecure ratio for both overlapping and non-overlapping domains and find no significant difference between them. 
Both groups exhibit high \insecure ratios; 48.4\% for overlapping domains (6,666 out of 13,762) and 53.6\% for non-overlapping domains (1,656 out of 3,087), respectively.

\paragraph{Insecure \https records and Name servers.}
However, when considering name servers, we found that \textit{domains served by \cf name servers exhibit a significantly higher \insecure ratio} compared to those that do not.
We find that 16,784 (99\%) domains are served by \cf name servers, and only 64 (1\%) domains are served by other entities' name servers (e.g., other hosting providers).\footnote{We are unable to retrieve the name server information for 11 domains.}
Specifically, domains using \cf name servers show a 49.5\% \insecure ratio, while those not using \cf name servers have a 14.1\% \insecure ratio, as shown in \Cref{tab:dnssec-stats}.
This notable discrepancy indicates that the high \insecure ratio of domains with \https records is primarily associated with domains using \cf name servers.

\begin{table}[t]
    \centering
    \resizebox{\columnwidth}{!}{%
    \begin{tabular}{c||r||c|c}
    \multirow{2}{*}{\bf Category} & \multicolumn{3}{c}{\bf Domains} \\
    & \bf Signed & \bf \texttt{Secure} & \bf \texttt{Insecure}  \\
    \hline
    \multicolumn{1}{l||}{without \https RR} & 46,850 & 35,688 (76.2\%) & 11,121 (23.7\%) \\
    \hline
    \multicolumn{1}{l||}{with \https RR} & 16,849 & 8,527 (50.6\%) & 8,322 (49.4\%)\\ 

    \multicolumn{1}{l||}{- \cf} & 16,784 & 8,471 (50.5\%) & 8,313 (49.5\%)\\ 
    \multicolumn{1}{l||}{- Non-\cf} & 64 & 55 (85.9\%) & 9 (14.1\%)\\ 
    \end{tabular}
    }
    \caption{The number of domains with signed records and their DNSSEC validation results, as of January 2nd, 2024. Domains with \https records are broken down based on their name servers (i.e., \cf or Non-\cf).
    We validate the DNSSEC chain of \https records if a domain published \https records, and the DNSSEC chain of \dnskey records for a domain without \https records. Note that \bogus validation results are omitted as there are no \bogus \https records.}
    \label{tab:dnssec-stats}
\vspace{-7mm}
\end{table}

Given the well-known issue that domains using a third-party DNS operator instead of their registrar's DNS service, often fail to upload necessary \ds records themselves~\cite{chung2017understanding},\footnote{Conversely, if a domain uses its registrar as the DNS operator, the registrar is capable of autonomously generating and uploading the domain's \ds records.} we further investigate the registrars of those domains.
We specifically examine the congruence between DNS operators and registrars for domains supporting DNSSEC.
To this end, we extract registrar information from additional Whois database searches.\footnote{We are able to gather Whois information for 88\% of domains with signed \https records.}
We then analyze this congruence for two domain groups based on \https record support: (i) signed domains \textit{without} \https records (i.e., the first row in \Cref{tab:dnssec-stats}) and (ii) signed domains \textit{with} \https records (i.e., the second row in \Cref{tab:dnssec-stats}).
Here, we simply determine congruence by checking whether a domain uses name servers known to be associated with a registrar; for instance, if a domain's registrar is \cf and it uses \texttt{amir.ns.cloudflare.com.} as a name server, this is classified as congruent.
First, we observe that the (i) DNSSEC-supporting domains that do not publish \https records have a 58\% alignment between their DNS operator and registrar.
Next, focusing on the (ii) DNSSEC-supporting domains with HTTPS records, we find that the top 10 popular registrars cover only 61.6\% of these domains. 
This indicates a varied distribution of registrars of these domains, particularly given that 99\% of the domains with \https records are served by \cf name servers.
Consequently, only 26\% of domains with signed \https records use the same DNS operator and registrar (among these 99\% use \cf for both services), which potentially accounts for their higher \insecure ratio.

\begin{figure}[!htbp]
\vspace{-3mm}
    \centering
    \setlength{\abovecaptionskip}{5pt}
    \setlength{\belowcaptionskip}{1pt}
    \includegraphics[width=.97\columnwidth]{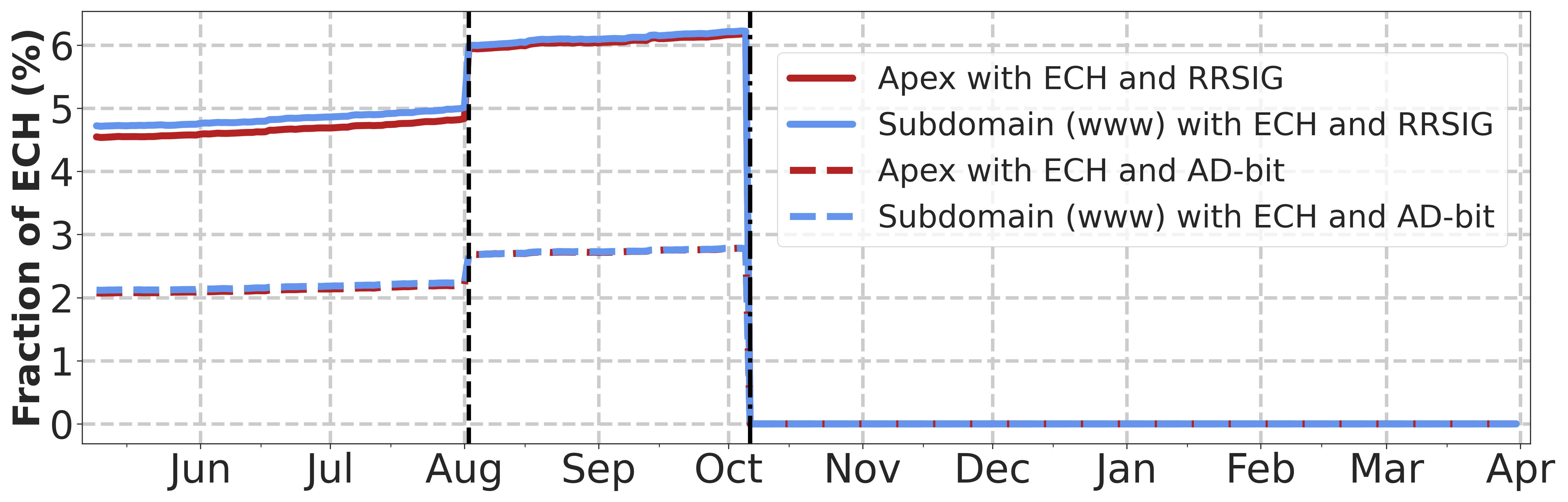}
    \caption{The percentage of overlapping domains with signed \https records and ECH parameter. The vertical dashed line (August 1st, 2023) denotes the source change of the Tranco list, and the vertical dash-dotted line (on Oct. 5, 2023) shows the date that \cf disabled ECH from its domains. }
    \label{fig:echrrsig}
\vspace{-3mm}
\end{figure}

\paragraph{ECH with DNSSEC trend.}
Before October 5th, 2023 (the date Cloudflare disabled ECH), less than 6\% of overlapping domains with HTTPS and ECH are signed, and only half of them can be validated, as shown in Figure~\ref{fig:echrrsig}. Note that the y-axis ticks represent percentages ranging from 0\% to 7\%. 